\documentclass[12pt,final]{article}
\usepackage{graphicx}
\usepackage{makecell}

\newcommand{\bq}{\begin{eqnarray}}
\newcommand{\bqn}{\begin{eqnarray*}} \newcommand{\enq}{\end{eqnarray}}
\newcommand{\enqn}{\end{eqnarray*}}

\usepackage{bm, amsmath,latexsym,amsthm,amssymb,amsbsy,amsfonts,longtable,lscape,ctable, enumerate, threeparttable,subfigure, mathtools}
\usepackage{algorithm, algorithmic}
\usepackage{natbib}
\usepackage{bibunits}
\usepackage[titletoc,toc,title]{appendix}

\newcommand*{\tran}{^{\mkern-1.5mu\mathsf{T}}}

\newtheorem{remark}{Remark}

\newtheorem{assumption}{Assumption}
\newtheorem{theorem}{Theorem}

\newtheorem{definition}{Definition}
\DeclareMathOperator*{\argmin}{arg\, min} 
\DeclareMathOperator\erf{erf}

\renewcommand\vec{\mathbf}
\renewcommand{\matrix}[1]{\mathbf{#1}}
\usepackage[final]{hyperref}
\hypersetup{
  colorlinks,%
  citecolor=blue,%
  filecolor=black,%
  linkcolor=blue,%
  urlcolor=black
}

\title{Unified Robust Estimation}

\author{Zhu Wang\\
  The University of Tennessee Health Science Center\\
Department of Preventive Medicine\\
Division of Biostatistics\\
66 North Pauline Street\\
Memphis, TN 38163\\
  E-mail: zwang145@uthsc.edu}
\begin{document}
 
 \maketitle
 \vspace{-0.5in}

\begin{abstract}
    Robust estimation is primarily concerned with providing reliable parameter estimates in the presence of outliers. Numerous robust loss functions have been proposed in regression and classification, along with various computing algorithms. In modern penalised generalised linear models (GLM), however, there is limited research on robust estimation that can provide weights to determine the outlier status of the observations. This article proposes a unified framework based on a large family of loss functions, a composite of concave and convex functions (CC-family). Properties of the CC-family are investigated, and CC-estimation is innovatively conducted via the iteratively reweighted convex optimisation (IRCO), which is a generalisation of the iteratively reweighted least squares in robust linear regression. For robust GLM, the IRCO becomes the iteratively reweighted GLM. The unified framework contains penalised estimation and robust support vector machine and is demonstrated with a variety of data applications.
\end{abstract}
\begin{bibunit}[apalike]
\section{Introduction}
Outliers are a small proportion of observations that deviate from the majority and can substantially cause bias in standard estimation methods. 
This problem has been tackled by robust estimation, which has a long history in statistical methodology research and applications \citep{hampel1986robust, maronna2019robust, heritier2009robust}. Denote response variables $y_i$, 
a $(p+1)$-dimensional predictor $\vec x_i=(x_{i0}, ..., x_{ip})\tran$ with the first entry 1, $i=1, ..., n$, and a $(p+1)$-dimensional coefficient vector
$\bm\beta=(\beta_0, \beta_1, ..., \beta_p)\tran$. Robust estimation can be achieved by minimising a loss function
\begin{equation}\label{eqn:Gamma}
  \argmin_{\bm\beta}\sum_{i=1}^n\Gamma(y_i, \vec x_i\tran\bm\beta),
\end{equation}
where popular choice of $\Gamma$ in linear regression is the Huber loss, Andrews loss or Tukey's biweight loss. The numerical solutions are typically computed through the so-called iteratively reweighted least squares (IRLS): 
\begin{equation}\label{eqn:wls}
  \argmin_{\bm\beta}\sum_{i=1}^n w_i(y_i - \vec x_i\tran\bm\beta)^2,
\end{equation}
where weights $w_i$ depend on the loss function $\Gamma$ such that smaller weights are assigned to those observations with larger residuals in magnitude. That is, outliers receive smaller weights. The weights should be understood as $w_i(y_i, \vec x_i, \vec x_i\tran\bm\beta)$ in general.
The M-estimators, however, can be defined directly using optimisation problem ~(\ref{eqn:wls}) without the need to introduce the minimisation problem~(\ref{eqn:Gamma}).
\subsection{Robust logistic regression}\label{sec:rlr}
For binary outcomes $y_i\in\{0, 1\}$, a robust logistic regression can be obtained by three approaches. First, the parameters can be estimated by a weighted maximum likelihood estimation (WMLE) or equivalently, a weighted minimum negative likelihood estimation 
\begin{equation}\label{eqn:wlog}
    \begin{aligned}
        &\argmin_{\bm\beta}\sum_{i=1}^n w_i s(\vec x_i\tran\bm\beta, y_i)\\
        &s(\vec x_i\tran\bm\beta, y_i)=-\left(y_i\log p_i(\bm\beta) + (1-y_i)\log(1-p_i(\bm\beta)\right)\\
        &p_i(\bm\beta)=\textrm{Pr}(y_i=1|\vec x_i, \bm\beta)=\frac{\exp(\vec x_i\tran\bm\beta)}{1+\exp(\vec x_i\tran\bm\beta)}.
\end{aligned}
\end{equation}
The weights $w_i$ include functions of the deviance and functions of predictors \citep{green1984iteratively, carroll1993robustness}. A modified method is a weighted estimation equation with a bias correction for consistent estimator \citep{heritier2009robust}. 

Second, \citet{pregibon1982resistant} proposed a composite loss function approach given by 
\begin{equation}\label{eqn:rlog1}
  \argmin_{\bm\beta}\sum_{i=1}^n g(s(\vec x_i\tran\bm\beta, y_i)),
\end{equation}
where $g$ is a strictly increasing Huber type function. This estimator was designed
to give less weight to observations poorly fitted by the model. 
Other functions $g$ have been proposed in \citet{bianco1996robust}, although the estimators may not exist in some applications. 
To address this issue, \citet{croux2003implementing} proposed different $g$ functions along with a somewhat complex algorithm. 

Third, 
with a focus on prediction, estimation can be achieved
by optimising a robust logistic loss function.
\citet{park2011robust, wang2018robust} 
have developed computing algorithms for truncated logistic loss functions, which are Fisher-consistent in classification, meaning that the population minimiser of the loss function leads to the Bayes optimal rule of classification \citep{lin2004note}. 
However, unlike traditional M-estimation, these approaches fail to retain the weights as a useful diagnostic for the outlier status of the observations.

If the analysis prioritises robust prediction, 
a natural generalisation of robust logistic regression is sought. An ideal estimation approach should fulfil four criteria:
     \begin{enumerate}[i.]
    \item The estimator should be obtained from a loss function satisfying Fisher consistency, which is a fundamental issue from the statistical learning perspective. 
    \item A shrinkage estimator can be derived by optimising a penalised loss function.
        Penalised estimation can improve prediction accuracy and simultaneously conduct parameter estimation and variable selection
             \citep{Tibs:1996, fan2001variable}. 
    \item The estimation should generate weights to indicate the outlier status of the observations. 
    \item The estimator should be computable using a reliable computer algorithm, and it would be advantageous if the algorithm can be generalised to other robust estimation problems.
     \end{enumerate}
However, previous robust logistic regression methods only satisfy some of the criteria but not all of them.
\subsection{Contribution}
We present a novel and unified approach to robust logistic estimation that fulfils all the requirements of the ideal approach mentioned earlier. Our method extends to robust generalised linear models (GLM) and other related problems, offering a versatile solution.
Our contributions can be summarised as follows:

First, we introduce a unified family of robust loss functions, which is a composite of concave and convex functions, known as the CC-family. This family encompasses well-known classical robust loss functions in statistics and data science, such as Huber loss, Andrews loss, biweight loss, robust logistic, and hinge loss. Moreover, it also includes a novel robust exponential family.

Second, we propose a new estimation framework that optimises the loss functions within the CC-family. The parameters are estimated using the iteratively reweighted convex optimisation (IRCO) technique, which is a generalisation of the iteratively reweighted least squares (IRLS) used in robust linear regression. The estimated weights provide valuable insights into the outlier status of observations. Additionally, we extend the IRCO method to handle penalised estimation.

Overall, our approach unifies various robust estimation techniques and offers a flexible and efficient solution for various statistical problems.


\subsection{Related work}

The CC-family encompasses various robust loss functions found in the literature. The concave $g$ functions within the CC-family include Huber, Andrews, and biweight type functions. In the context of robust logistic regression, the CC-family comprises Huber's type $g$ function from \citet{pregibon1982resistant} and a truncated $g$ function from \citet{bianco1996robust}. Additionally, a rescaled hinge loss \citep{xu2017robust} also belongs to the CC-family. Notably, the IRCO incorporates the IRLS as a special case for robust linear regression. Moreover, for specific members of the CC-family, the IRCO can be slightly modified to conduct least trimmed squares estimation, and the iteratively reweighted support vector machine in \citet{xu2017robust} represents a special case of the IRCO. It's worth mentioning that the IRCO offers two approaches for computing weights, with one being simpler than the approach used in \citet{xu2017robust}.

Alternatively, there is another algorithm for the truncated hinge loss, known as the difference-of-convex (DC) algorithm \citep{wu2007robust}. The DC algorithm decomposes the loss function $\Gamma$ into a difference of two convex functions, whereas the IRCO involves a composite of convex and concave functions. However, the DC algorithm does not update observation weights corresponding to the outlier status, and most CC-family members do not have a simple DC formula except for the truncated loss.

The requirement for a concave function $g$ in the CC-family offers several benefits. 
For instance, while a composite gradient descent approach can be easily developed to solve a more general composite algorithm and provide greater flexibility in solutions, this algorithm lacks the weights as a distinctive characteristic of the outlier status of observations.
Moreover, a gradient method may not be the best option in certain scenarios, such as when dealing with the robust hinge loss for support vector machines (SVM) with nonlinear kernels like the Gaussian kernel. In contrast, the IRCO for the robust hinge loss effectively corresponds to the iteratively reweighted SVM and can be conveniently implemented using existing software.

The remainder of this article is structured as follows. In Section 2, we present the structure and characteristics of the CC-family. Section 3 details the IRCO for the CC-estimators, explores its convergence properties, and establishes its connections with other algorithms. In Section 4, we illustrate the extensive applications of CC-estimators using both simulated and real data. We showcase a variety of CC-estimators in robust estimation tasks, including regression and GLMs with penalised estimation.
In Section 5, we conclude the article with further discussions. The online Supplementary Information provides additional applications, such as the robust SVM, and includes technical proofs.

\section{Composite loss functions}\label{sec:robust}
 
\begin{figure}[!htbp]
 \begin{center}
   \includegraphics[scale=0.8]{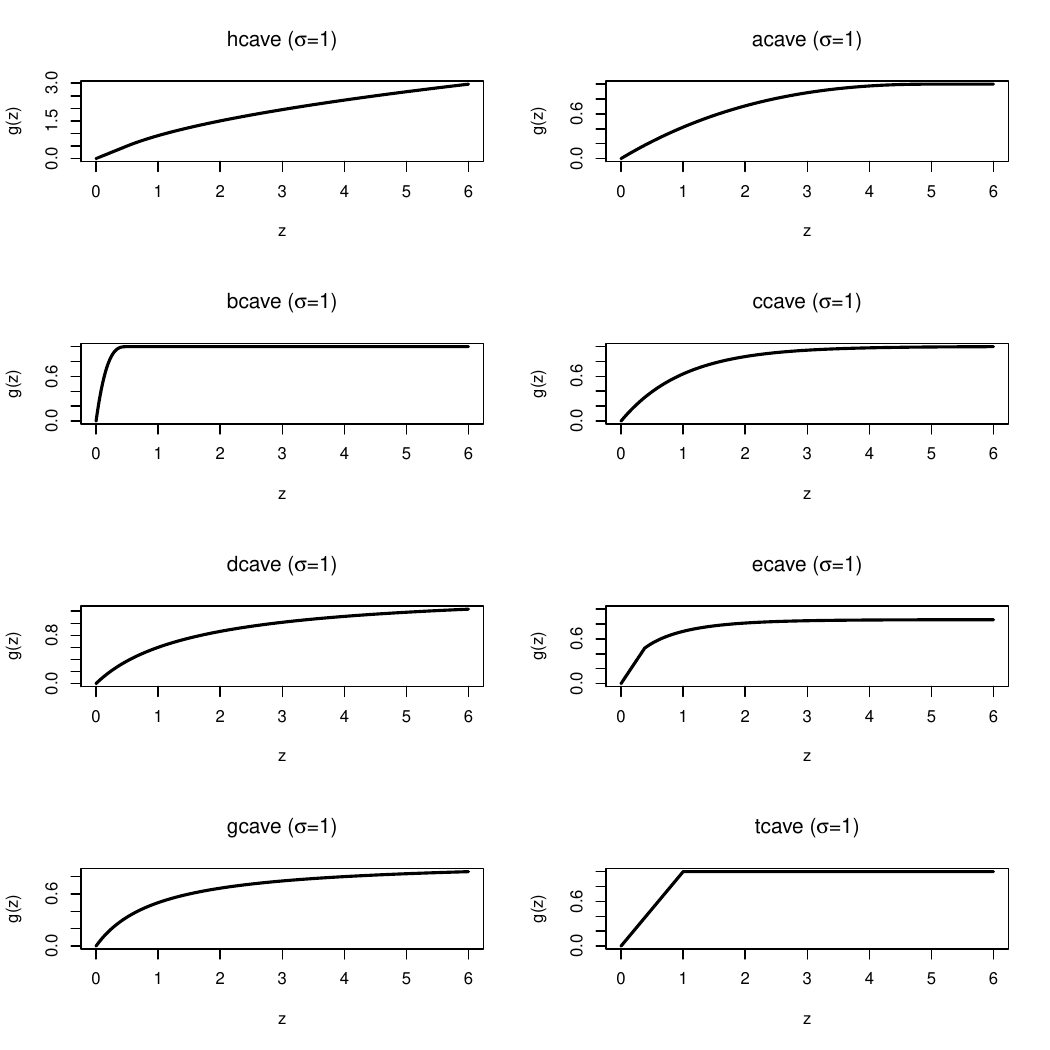}
 \caption{Concave component.}
    \label{fig:cfung}
  \end{center}
\end{figure}

The literature has extensively explored a variety of robust loss functions, which are documented in Table~\ref{tab:tab1} \citep{maronna2019robust, xu2017robust,wang2018robust,         wang2022mm}. 
These functions can be organised as composite functions, forming the basis of the concave-convex (CC) family.
 \begin{definition}[CC-family]\label{assu:cc}
    The CC-family contains composite functions $\Gamma=g\circ s$ satisfying the following conditions:
     \begin{enumerate}[(i)]
     \item $g$ is a nondecreasing closed concave function whose domain is the range of function $s$
     \item $s$ is convex on $\mathbb{R}$. 
 \end{enumerate}  
 \end{definition}
The $g$ component, which is concave, robustifies the classical nonrobust estimator obtained from the convex $s$ component, such as least squares and negative likelihood functions. The concave property of $g$ is necessary for the IRCO algorithm. Table~\ref{tab:gs} provides a list of concave components derived from Table~\ref{tab:tab1}. 
Some modifications are required to convert the $g$ of Qloss in Table~\ref{tab:tab1} to ecave, ensuring that the latter is concave with a bounded and continuous derivative. The ecave function is related to erf, the Gaussian error function. Similarly, gcave is constructed from the $g$ of Gloss, ensuring its derivative is bounded and continuous. 
As shown in Figure~\ref{fig:cfung}, all functions, except for hcave, are bounded.

The concave component, along with the derived composite function, is parameterised by $\sigma$, which controls the robustness of the estimation. A smaller value of $\sigma$ allows for more robust estimation. The role of parameter $\sigma$ has been extensively studied in the literature \citep{maronna2019robust, wu2007robust}. The IRCO algorithm in Section~\ref{sec:robest} will shed light on the impact of $\sigma$ on the estimation process.

Table~\ref{tab:dfun} presents the convex components, which serve as fundamental building blocks in various data analysis theories and applications. For regression problems, the convex component can be Gaussian or $\epsilon$-intensive, which is a crucial device for support vector machine regression \citep{Hast:esl:2009}.
In classification tasks, convex components can be derived from GaussianC, binomial, or hinge loss functions. The GLMs are obtained from the exponential family.

For convenience, Gaussian and binomial losses are separated from the exponential family. In the exponential family, $s(u)$ represents the negative log-likelihood function for certain functions $a(\cdot), b(\cdot), \text{ and }c(\cdot)$. It is well known that the cumulant function $b(\cdot)$ is convex in its domain \citep[Prop. 3.1]{wainwright2008graphical}.
Indeed, $s(u)$ is convex in the exponential family. However, it is important to note that $s(u)$ can be negative in certain cases. To construct a valid composite function $g\circ s$ when the domain of $g$ is non-negative, one can make the substitution $s(u)$ with $s(u)-C(y)$, where $C(y)$ is data-dependent and chosen such that $s(u)-C(y) \geq 0$. This can be achieved since $s(u)$ is minimised when $u$ is equivalent to $y$ via a link function in the exponential family. This modification ensures that the composite function remains valid and satisfies the non-negativity constraint of $g$.

Furthermore, by employing common operations with convex functions, it is possible to obtain new members of the CC-family. The corresponding subdifferentials of these functions can be particularly useful in the IRCO algorithm.
\begin{theorem}\label{thm:linear}
    Let $\Gamma_1=g_1\circ s$ and $\Gamma_2=g_2\circ s$ be members of the CC-family $\Omega$ and $c_1, c_2 \geq 0, g=c_1 g_1+c_2 g_2$.  
    Then $\Gamma=g\circ s \in \Omega$ holds and 
    \begin{equation}\label{eqn:linear}
        \partial(-g(z))=c_1\partial(-g_1(z))+c_2\partial(-g_2(z))
    \end{equation}
    for any $z$ from int (dom $g$)=int (dom $g_1)\cap$int (dom $g_2$), where int (dom $g$) is the interior of domain of $g$.
\end{theorem}

\begin{theorem}\label{thm:max}
    Let $\Gamma_i=g_i\circ s, i=1, ...,m,$  be members of the CC-family $\Omega, g=\min_{1\leq i\leq m}g_i$. 
    Then $\Gamma=
    g\circ s \in \Omega$ holds. For any $z\in$ int (dom $g$)=$\cap_{i=1}^m$ int (dom $g_i$), we have 
    \begin{equation}\label{eqn:max}
        \partial(-g(z))=\text{Conv}\{\partial(-g_i(z))|i \in I(z)\},
    \end{equation}
where
    \begin{equation*}
    \begin{aligned}
        \text{Conv}\{x_1,..., x_m\}& = \left\{x = \sum_{i=1}^m a_ix_i | a_i \geq 0, \sum_{i=1}^m a_i= 1\right\},\\
                                I(z)&=\{i: g_i(z)=g(z)\}.
    \end{aligned}
    \end{equation*}
\end{theorem}
The following properties characterise the robustness of loss functions and are also closely related to the IRCO algorithm.
\begin{theorem}\label{thm:concave11}
  Assume that $g:\text{range of }s \to \mathbb{R}$, where $\text{range of } s$ is open, $g$ and $s$ are twice differentiable, $s'(u) \neq 0$.
  Then $g$ is concave if and only if for every $ u \in \text{dom }s$, the following holds:
  \begin{equation}\label{eqn:lemconc0}
    \frac{s''(u)}{s'(u)}\Gamma'(u) \geq  \Gamma''(u).
  \end{equation}
  \end{theorem}
  For convex function $s$, since $s''(u) \geq 0$, (\ref{eqn:lemconc0}) is equivalent to
  \begin{equation*}\label{eqn:lemconc00}
    \frac{\Gamma'(u)}{s'(u)} \geq  \frac{\Gamma''(u)}{s''(u)},
  \end{equation*}
  provided that $s''(u) \neq 0$. 
  For instance, with $s(u)=u^2/2$, we have for every $u$,
    \begin{equation*}\label{eqn:conc1}
    \frac{\Gamma'(u)}{u} \geq \Gamma''(u).
  \end{equation*}
  Note that $\frac{\Gamma'(u)}{u}$ is the weight used for M-estimator in robust estimation \citep{maronna2019robust}. Likewise, $g'(s(u))=\Gamma'(u)/{s'(u)} $ is the weight in the IRCO.
  
 Theorem~\ref{thm:concave11} is related to the absolute risk aversion for function $s(u), u \geq 0$:
  \begin{equation*}
    ARA(u)=-\frac{s''(u)}{s'(u)}.
  \end{equation*}
  ARA is a popular metric in economics for utility function $s(u)$ that measures preferences over a set of goods and services \citep{pratt1964risk}. Assuming nondecreasing function $s$, we get $\Gamma'(u)=g'(s(u))s'(u) \geq 0$ for concave function $g$. Theorem~\ref{thm:concave11} implies that
  \begin{equation*}\label{eqn:con3}
    -\frac{s''(u)}{s'(u)} \leq  -\frac{\Gamma''(u)}{\Gamma'(u)}
  \end{equation*}
  for $\Gamma'(u) \neq 0$. Hence, $\Gamma(u)$ shows globally more risk averse than $s(u)$ if and only if $\Gamma(u)$ is a concave transform of $s(u)$.
 
  Theorem~\ref{thm:concave11} is applicable to many functions in the CC-family, for instance, concave component acave-dcave and gcave ($\sigma \geq 1)$, and convex component exponential family. The Huber's type $g$, however, is only piecewisely twice differentiable. In this case, the following similar results hold.
 \begin{theorem}\label{thm:concave2}
  Assume that $g: \text{range of }s \to \mathbb{R}$ is continuous, range of $s=(a, b)$, there is a subdivision $z_0=a < z_1 < ... < z_k=b$ of (a, b), $g$ is twice continuously differentiable on each subinterval $(z_{i-1}, z_i), i=1, ..., k$, $g$ has one-sided derivatives at $z_1, ..., z_{k-1}$ satisfying $D_-g(z_i) \leq D_+g(z_i)$ for $i=1, ..., k-1$, $s$ is twice differentiable, $s'(u)\neq 0$. Then $g$ is concave if and only if
  \begin{equation*}\label{eqn:con2}
    \frac{s''(u)}{s'(u)}\Gamma'(u) \geq  \Gamma''(u)
  \end{equation*}
  holds on each subinterval $(z_{i-1}, z_i), i=1, ..., k$.
  \end{theorem}
  Theorem~\ref{thm:concave2} is applicable to the CC-family with concave component hcave, ecave and gcave ($\sigma < 1$), and convex component exponential family. With $s(u)=u^2/2, u \geq 0$, $s$ is nondecreasing. The Gaussian induced loss functions have larger ARA than that of Gaussian, provided the ARA exists. For the Huber loss with concave component hcave, simple algebra shows that:
  \begin{equation*}\label{eqn:huber2_3}
  \begin{aligned}
    -\frac{s''(u)}{s'(u)} &=  -\frac{\Gamma''(u)}{\Gamma'(u)}, \text{ if } 0 < u < \sigma,\\ 
    -\frac{s''(u)}{s'(u)} &<  -\frac{\Gamma''(u)}{\Gamma'(u)}, \text{ if } u > \sigma. 
      \end{aligned}
  \end{equation*}
ARA is overlapped with the Gaussian loss when $0 < u < \sigma$ and greater than the Gaussian when $u > \sigma$. In other words, we obtain the well-known result: the Huber loss is the same as the Gaussian when $0 < u < \sigma$ and more robust than the Gaussian otherwise. 

Since hinge-type losses do not satisfy a piecewise twice differentiable assumption on the whole        domain, Theorem~\ref{thm:concave11} and \ref{thm:concave2} are not applicable.
\subsection{Regression}
\begin{figure}[!htbp]
  \begin{center}
    \includegraphics[scale=0.8]{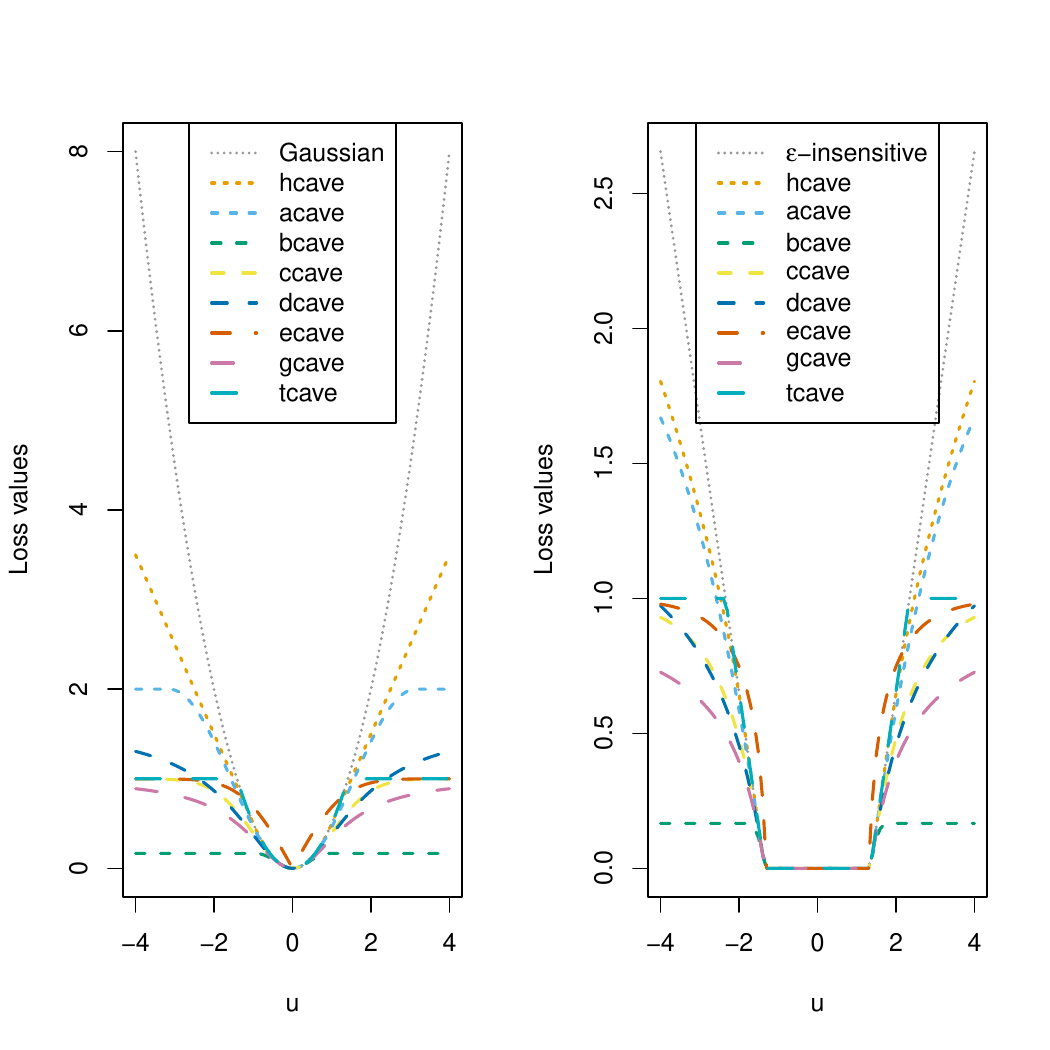}
 \caption{Convex component Gaussian, $\epsilon$-insensitive and their induced composite loss functions.}
    \label{fig:loss2}
  \end{center}
\end{figure}

\begin{figure}[!htbp]
  \begin{center}
    \includegraphics[scale=0.8]{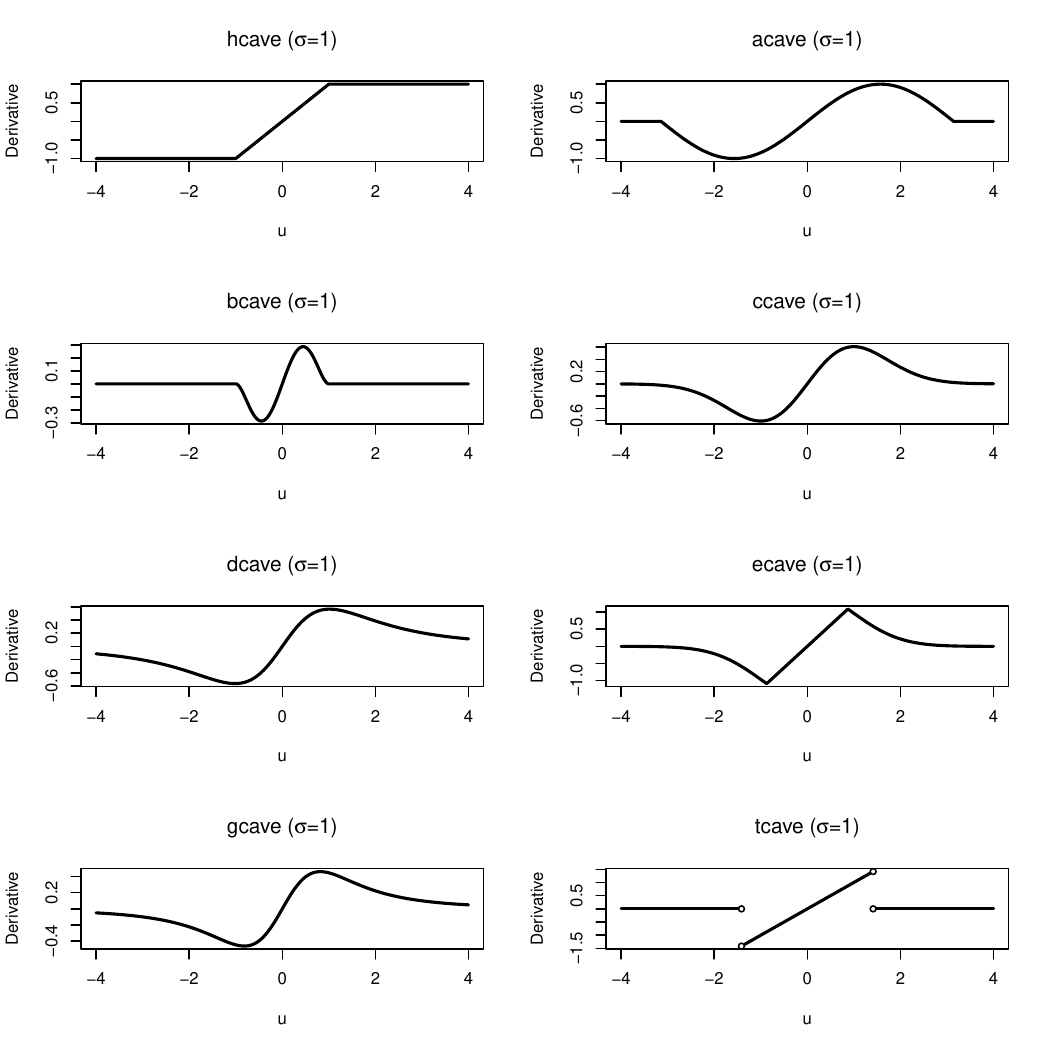}
 \caption{Derivatives of Gaussian induced composite loss functions.}
    \label{fig:cdderi}
  \end{center}
\end{figure}

The CC-family contains Gaussian-induced composite functions, as shown in Figure~\ref{fig:loss2}. In addition to classic robust loss functions, new members are introduced from dcave, ecave, and gcave. Figure~\ref{fig:loss2} also includes innovative $\epsilon$-insensitive induced loss functions. The composite functions are flatter than their convex counterparts and even become bounded except for hcave, making them more robust to outliers. The derivatives of Gaussian-induced loss functions are shown in Figure~\ref{fig:cdderi}. With monotone $\Gamma'$, the M-estimates can break down for high leverage outliers \citep[Section 5.3]{maronna2019robust}. However, except for hcave (Huber loss), all Gaussian-induced loss functions in Figure~\ref{fig:cdderi} are robust to high leverage outliers.
\subsection{Classification}
\begin{figure}[!htbp]
  \begin{center}
    \includegraphics[scale=0.8]{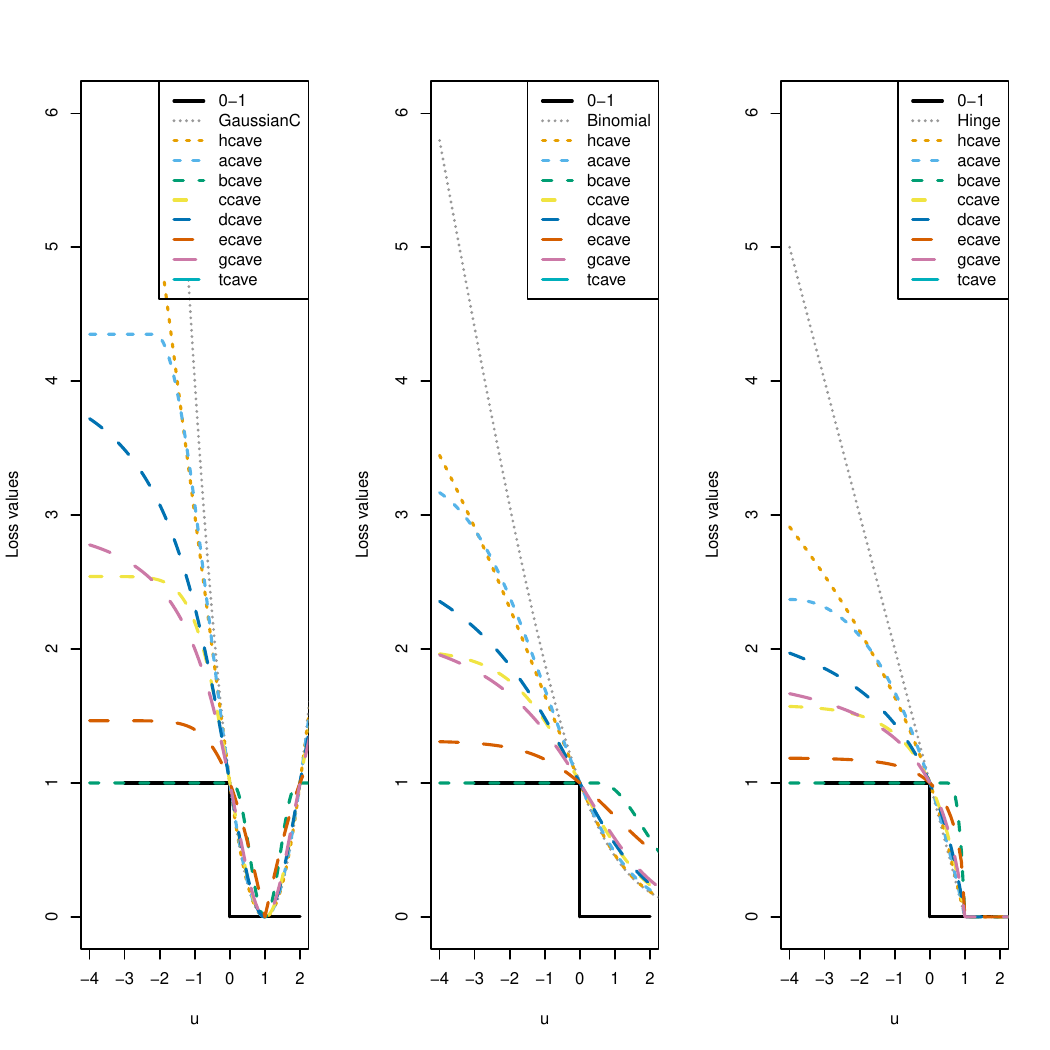}
 \caption{Convex component GaussianC, Binomial, Hinge loss and their induced composite loss functions.}
    \label{fig:loss3}
  \end{center}
\end{figure}

For a binary outcome $y$ taking values $+1$ and $-1$, the margin of a classifier $f$ is denoted by $u=yf$. Traditional classification problems utilise convex GaussianC, binomial, and hinge loss \citep{Hast:esl:2009}. These functions, along with their induced loss functions, are shown in Figure~\ref{fig:loss3}. The composite values are normalised such that $g(s(0))=1$, which effectively requires $\sigma \geq 1$ for tcave. The convex component loss functions are unbounded and cannot control
outliers well. On the other hand, the CC-family, except for hcave (Huber-type), is bounded, leading to more robust estimation.

The Fisher consistency of margin-based loss functions was initially studied in \citet{lin2004note}. In this article, we extend and present additional conditions for Fisher consistency:

\begin{enumerate}
    \item $ s(u)<s(-u), \ u>0$.
        \label{con1}
    \item $s'(0) < 0$.
        \label{con2}
    \item $g: \text{range of }s\to \mathbb{R}$ is strictly increasing.
        \label{con3}
    \item $g'(s(0))\neq 0$ exists.
        \label{con4}
    \item $g\circ s$ is a non-increasing function with $\sigma \geq 1$.
        \label{con6}
    \item If $\sigma=1$, then $1=g(s(0)) > g(s(1))$ and $g(s(0)) = g(s(-1))$ hold.
        \label{con5}
    \item If $\sigma > 1$, then $g'(s(0))\neq 0$ exists. 
        \label{con44}
\end{enumerate}
\begin{theorem}\label{thm:fisher}
    Assume that $\Gamma=g\circ s$.
Then for $Y \in \{-1, 1\},\Gamma(Yf(X))$ is Fisher-consistent if either of the following two sets of conditions holds:
  \begin{enumerate}[(i)]
      \item Conditions~\ref{con1}--\ref{con4} hold.
      \item Conditions~\ref{con2}, \ref{con6}--\ref{con44} hold.
  \end{enumerate}
\end{theorem}
Conditions 1 and 2 ensure that the function $s$ is Fisher consistent \citep{lin2004note}. 
 Case (ii) generalises the truncated hinge and logistic loss functions with $g=\min(\sigma, z)$ \citep{wu2007robust, park2011robust}.
Theorem \ref{thm:fisher} guarantees that many classification loss functions in the CC-family satisfy the Fisher consistency property. However, one exception is the composite of concave tcave and convex GaussianC. This composite function does not satisfy condition~\ref{con6}.

\section{Robust estimation}\label{sec:robest}
In this section, we present an overview of the estimation problem in the CC-family. We then discuss two different approaches in algorithm design for solving this estimation problem. Next, we provide a detailed description of the IRCO and its convergence results. Finally, we establish connections between the IRCO and the trimmed estimation method.
\subsection{Estimation problem}
Consider data-dependent convex component $s(u_i)$ given in Table~\ref{tab:dfun}, where
\begin{equation}\label{eqn:ui}
  u_i=
  \begin{cases}
    y_i-f_i, &\text{ for regression,}\\
    y_i f_i,  &\text{ for classification with $y_i \in [-1, 1]$},\\
    f_i,      &\text{ for exponential family.}
  \end{cases}
\end{equation}
Here $u_i$ may be seen as $u_i=u_i(\bm\beta)$ and $f_i=\vec x_i\tran\bm\beta$. 
Note that $u_i$ is linked to the linear predictor $f_i$ via (\ref{eqn:ui}), although more complex transformations may be used, such as in the case of nonlinear kernels of SVM. A CC-estimator is obtained by finding a solution that minimises the empirical loss
$L(\bm\beta)$ given by
\begin{equation}\label{eqn:mloss2}
    L(\bm\beta)=\frac{1}{n}\sum_{i=1}^n\Gamma(u_i(\bm\beta))=
    \frac{1}{n}\sum_{i=1}^n g(s(u_i(\bm\beta))).
\end{equation}
For logistic regression with $y_i \in \{0, 1\}$, we have
\begin{equation*}
    s(u_i)=-y_i\vec{x_i}\tran\bm\beta+\log(1+\exp(\vec{x_i}\tran\bm\beta)),
\end{equation*}
which is equivalent to the binomial loss in Table~\ref{tab:dfun} with the margin $u_i=y_i\vec{x_i}\tran\bm\beta, y_i \in [-1, 1]$. Another example is the Poisson regression:
\begin{equation*}
    s(u_i)=-y_i\vec{x_i}\tran\bm\beta+\exp(\vec{x_i}\tran\bm\beta).
\end{equation*}
In many applications, we optimise a penalised loss function $F: \mathbb{R}^{p+1} \to \mathbb{R}$:
	\begin{equation}\label{eqn:plik}
  F(\bm\beta)=
  L(\bm\beta)+ 
	\Lambda(\bm\beta),
\end{equation}
     where 
     \begin{equation*}
	     \Lambda(\bm\beta)= \sum_{j=1}^{p} \left(\alpha 
         p_\lambda(|\beta_j|) + \lambda\frac{1-\alpha}{2}\beta_j^2\right),
     \end{equation*}
$0 \leq \alpha \leq 1,
\lambda \geq 0$, and $p_\lambda(|\beta_j|)$ is the penalty function such as the LASSO \citep{Tibs:1996} or SCAD \citep{fan2001variable}.
Minimising the penalised loss function can avoid overfitting, provide shrinkage estimates and conduct variable selection. The loss function (\ref{eqn:mloss2}) is a special case of (\ref{eqn:plik}) with $\Lambda(\bm\beta)=0$, i.e., $\lambda=0$. 
\subsection{Algorithm design by the first-order condition of convexity}\label{sec:foc}
Suppose $h$ is a differentiable convex function on its convex domain. Function $h$, or equivalently, concave function $g=-h$ has the first-order condition for every $u, \hat u \in \text{dom }g$
\begin{equation}
    g(u)\leq g(\hat u) + g'(\hat u)(u-\hat u).
\end{equation}
Replace $u$ with $s(u)$, $\hat u$ with $s(\hat u)$. Thus we have 
\begin{equation}\label{eqn:foc3}
    g(s(u)) \leq g(s(\hat u)) + g'(s(\hat u))(s(u)-s(\hat u)) = \gamma(u|\hat u).
\end{equation}
Then $\gamma(u|\hat u)$ majorises $\Gamma(u)=g(s(u))$ at $\hat u$ because we have for every $u$
\begin{equation}\label{eqn:foc5}
	\Gamma(u) \leq \gamma(u|\hat u), \ 
	\Gamma(\hat u) = \gamma(\hat u|\hat u).
\end{equation}
For a nondifferentiable function $g$, similar results hold if the derivative in the first-order condition is replaced with the subgradient. The algorithm follows the majorisation-minimisation (MM) framework \citep{lange2016mm}, which is an iterative procedure. Given an estimate $u^{(k)}$ in the $k$th iteration, $\gamma(u|u^{(k)})$ is minimised at the $k+1$ iteration to     obtain an updated minimiser $u^{(k+1)}$.
 This     process is repeated until convergence.
 The MM algorithm generates a descent sequence of           estimates:
 \begin{equation}\label{eqn:mm}
   \Gamma(u^{(k+1)}) \leq 
     \gamma(u^{(k+1)}|u^{(k)}) \leq 
     \gamma(u^{(k)}|u^{(k)})  =  
   \Gamma(u^{(k)}). 
 \end{equation}
\subsection{Algorithm design by the Fenchel convex conjugate}\label{sec:fcc}
Let $\varphi$ be the convex or Fenchel conjugate of function $h$ defined by:
\begin{equation*}
  \varphi(v)=\sup_{z \in \text{dom } h} (zv-h(z)).
\end{equation*}
The conjugate $\varphi$ is convex on $\text{dom } \varphi$. And conjugate of $\varphi$ is restored if $h$ is a closed convex function \citep[Fenchel--Moreau theorem]{lange2016mm}:
\begin{equation*}\label{eqn:conj2}
\begin{aligned}
  h(z)=&\sup_{v \in \text{dom } \varphi} (zv-\varphi(v))\\
      =&-\inf_{v \in \text{dom }\varphi }(z(-v)+\varphi(v)).
      \end{aligned}
\end{equation*}
Let $h=-g$, where $g$ is concave. Thus we obtain 
\begin{equation*}
	g(z)=\inf_{v \in \text{dom } \varphi} (z(-v)+\varphi(v)).
\end{equation*}
With $z=s(u)$ we get
\begin{equation*}\label{eqn:mm3}
	g(s(u))=\inf_{v \in \text{dom } \varphi} (s(u)(-v)+\varphi(v)).
\end{equation*}
Define
\begin{equation}\label{eqn:coco1}
	\Gamma(u)=g(s(u)), \ \zeta(u, v)=s(u)(-v)+\varphi(v).
\end{equation}
Then $\zeta(u, v)$ majorises $\Gamma(u)$ at $\hat v, \text{ where}\ \hat v=\argmin_v s(u)(-v)+\varphi(v).$ 
An MM algorithm can be developed to minimise $\Gamma(u)$ via function $\zeta(u, v)$ in an alternating scheme. 
First, given the current value of $\hat u$, we solve $\hat v=\argmin_v s(\hat  u)(-v)+\varphi(v)$. Second, with the current value of $\hat v$, we minimise $\zeta(u, \hat v)$ with respect to $u$. This process repeats until convergence. Different from the first-order condition design in Section~\ref{sec:foc}, the Fenchel conjugate must be computed. Furthermore, a middle step is required to optimise the $\zeta(u, v)$ in each iteration. However, it will be formally proved in
Theorem~\ref{thm:conv1} that the two designs lead to the same solution.
\subsection{IRCO}
\begin{figure}[!htbp]
 \begin{center}
   \includegraphics[scale=0.8]{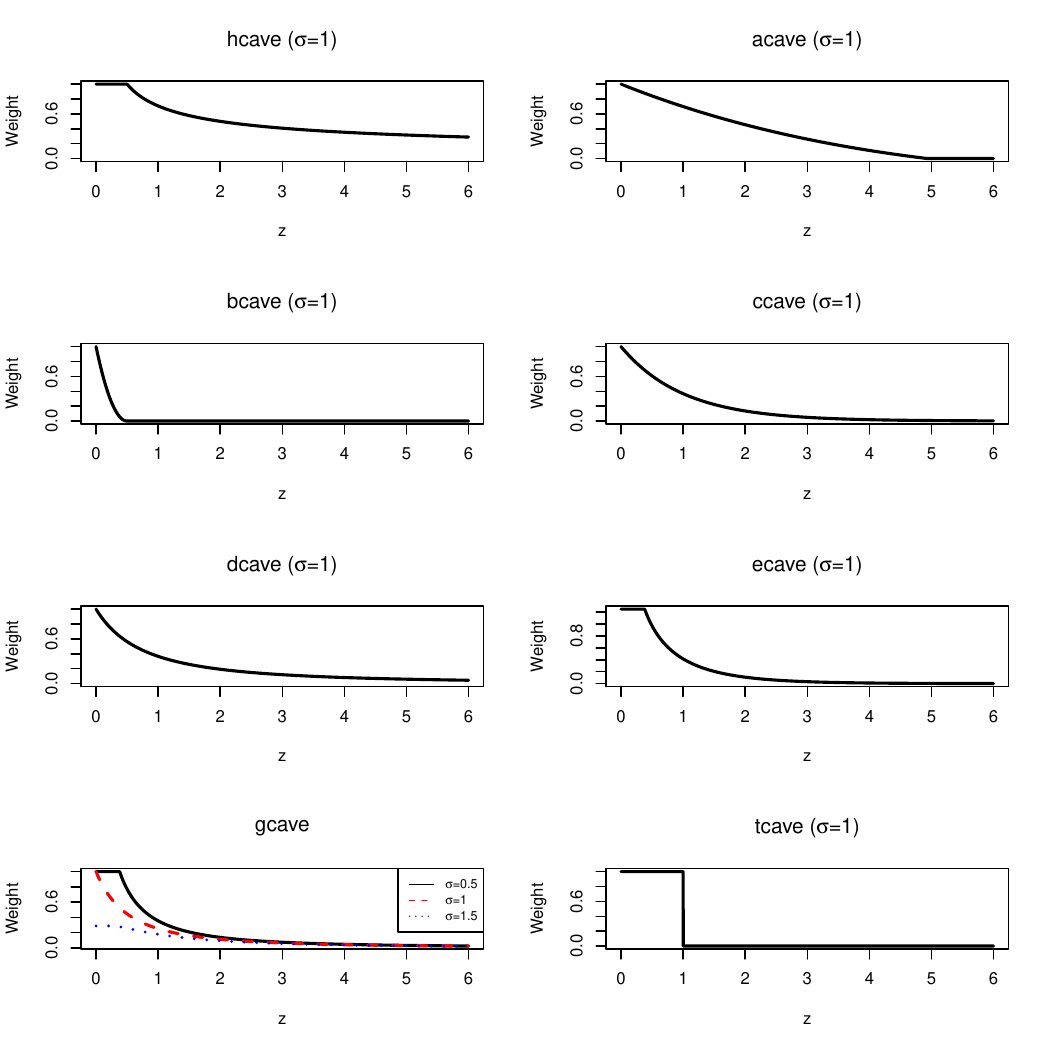}
 \caption{Weight function $-\partial(-g(z))$.} 
    \label{fig:cfunwt}
  \end{center}
\end{figure}

The IRCO to minimise data-driven loss $F(\bm \beta)$ in (\ref{eqn:plik}) is given in Algorithm~\ref{alg:pcoco}.
\begin{algorithm}[!htbp]
  \begin{algorithmic}[1]
    \caption{IRCO}\label{alg:pcoco}
    \STATE \textbf{Initialise} $\bm\beta^{(0)}$ and set $k=0$
    \REPEAT
    \STATE Compute $u_i(\bm\beta^{(k)})$ in (\ref{eqn:ui}) and $z_i=s(u_i(\bm\beta^{(k)})), i=1, ..., n$
    \STATE Compute $v_i^{(k+1)}$ via $v_i^{(k+1)}\in \partial(-g(z_i))$ or $z_i \in \partial \varphi(v_i^{(k+1)}), i=1, ..., n$ 
      \STATE Compute $\bm\beta^{(k+1)} = \argmin_{\bm\beta} \sum_{i=1}^n s(u_i(\bm\beta))(-v_i^{(k+1)})+\Lambda(\bm\beta)$
    \STATE $k= k+1$
    \UNTIL convergence of $\bm\beta^{(k)}$
  \end{algorithmic}
\end{algorithm}
\begin{remark}\label{rmk:step3}
    The two approaches to computing the weights in Step 4 correspond to the two algorithm designs in Section~\ref{sec:foc} and \ref{sec:fcc}.
    \citet{xu2017robust} took the approach in Section~\ref{sec:fcc} for the composite of the ccave and hinge loss. They derived $\varphi$ and its derivative to compute the weights. 
    For many applications, the approach in Section~\ref{sec:foc} is much simpler
    since no middle steps or derivations are required.
    Furthermore, the weights from the two approaches are the same, thanks to the Fenchel--Moreau theorem. See Theorem~\ref{thm:conv1} and its proof below. 
  \end{remark}
    \begin{remark}\label{rmk:exi}
	Step 4 assumes that $v_i^{(k+1)}$ exists. This can be justified as follows. 
If $v_i^{(k+1)}$ is an interior point of $\textup{dom } \varphi$, then $\partial \varphi(v_i^{(k+1)})$ is a nonempty bounded set 
        since conjugate function $\varphi$ is closed and convex \citep[Theorem 3.1.13]{nesterov2004introductory}. 
Likewise, if $-g$ is closed and convex, and $z_i$ is an interior point of $\textup{dom } g$, then $\partial(-g(z_i))$ is a nonempty bounded set. Care must be taken on the boundary points. Corresponding to dom $g=\{z: z \geq 0\}$ in Table~\ref{tab:gs}, 
on boundary point $z=0$, $g$ must be chosen such that $\partial(-g(z))$ is not empty or unbounded. For instance, ecave and gcave ($0 < \sigma < 1$) are piecewisely constructed to achieve bounded derivative at the origin. 
  For acave, while $g'(0)$ does not exist, it is simple to choose
  \begin{equation}\label{eqn:rmk1}
  g'(0)=\lim_{z \to 0+} g'(z). 
  \end{equation}
  \end{remark}
\begin{remark}\label{rmk:sigma}
  Step 5 amounts to a weighted minimisation problem with weights $-v_i^{(k+1)}$. Since $-g(z)$ is nonincreasing convex, we have $v_i^{(k+1)} \leq 0, i=1, ..., n$. Furthermore, $v_i^{(k+1)}$ is a nondecreasing function of $z_i$. 
	See Table~\ref{tab:sub} and Figure~\ref{fig:cfunwt}. 
    Thereby, `clean data' with small values of $z_i$ will receive larger weights, while outliers with a large value of $z_i$ will receive smaller weights. 
    Note $\sigma$ is suppressed in $g(z)$. For hcave, acave, bcave, ccave and tcave, we obtain $\partial (-g(z,\sigma)) \to -1$ as $\sigma \to \infty$. While a subdifferential is a set by definition, to simplify notations, we interchange between set $\{A\}$ and $A$ when $A$ is the sole element in the set. The relationship between robustness and weights in Table~\ref{tab:sub} suggests that a larger value $\sigma$ is less robust. Therefore, one may tune the $\sigma$ value from a large value to a
    small value, that is, from a classical estimator to a robust estimator and select an optimal value of $\sigma$ according to some data-driven criteria. We adopt this procedure in Section 4. 
\end{remark}
\begin{remark}
  The IRCO is a generalisation of the IRLS to compute M-estimators \citep[section 4.5.2]{maronna2019robust}.
    For $\Gamma(u)=g(z), z=s(u)=u^2/2$, at the $k$-th iteration of the IRLS, we compute
  \begin{equation*}\label{eqn:irwls}
    \argmin\sum_{i=1}^n w_{i}^{(k+1)}(u_i)u_i^2,
  \end{equation*}
  where the weights are defined by
  \begin{equation}\label{eqn:weight}
    w_i^{(k+1)}(u_i)=
    \begin{cases}
    \Gamma'(u_i)/u_i &\text{ if }u_i \neq 0,\\
    \Gamma''(0)       &\text{ if }u_i = 0.
  \end{cases}
  \end{equation}
    It can be shown that $w_i^{(k+1)}(u_i)=-\partial(-g(z_i))$ if $g$ is differentiable at $z_i=s(u_i)$ since we have:
    \begin{equation}
        -\partial(-g(z_i))=g'(z_i)=g'(s(u_i))=\frac{\Gamma'(u_i)}{s'(u_i)}=\frac{\Gamma'(u_i)}{u_i}.
    \end{equation} 
    The remedy in (\ref{eqn:weight}) for $u_i=0$ is the same as (\ref{eqn:rmk1}).
\end{remark}
\begin{remark}
Step 5 involves a penalised estimation problem, and we utilise an efficient coordinate descent algorithm, as described in \citet{Frie:regu:2010}. In nonconvex optimisation, the IRCO typically seeks a local solution, and it is possible to obtain different local solutions with different initial values. Hence, the algorithm may begin with various initial values and determine the best solutions afterwards. For the numerical study in Section 4, we simply initialise $\bm\beta^{(0)}=\vec{0}$, and the simulation and data analysis results support this choice.
\end{remark}
We have obtained convergence results for the IRCO, and the penalty assumptions are provided in the Appendix in the Supplementary Information.
\begin{theorem}\label{thm:conv1}
  Suppose that $g$ is a concave component in the CC-family, and $g$ is bounded below. 
	\begin{enumerate}[(i)]
		\item The loss function values $F(\bm\beta^{(k)})$
  generated by Algorithm~\ref{alg:pcoco} are nonincreasing and converge. 
	\item Assume that $g$ and $s$ are differentiable,  	$\zeta(u, v)=s(u)(-v)+\varphi(v)$ is jointly continuous in $(u, v)$, $\varphi$ is the conjugate function of $-g$,
        $\nabla L(\bm\beta)=\nabla\ell(\bm\beta|\bm\beta^{(k)}),$
  where the surrogate loss is given by
\begin{equation*}
	\ell(\bm\beta|\bm\beta^{(k)})=\sum_{i=1}^n \zeta(u(\bm\beta), v(\bm\beta^{(k)})),
\end{equation*}
 and $p_\lambda(|\cdot|)$ satisfies mild assumptions. 
			Then every limit point of the iterates generated by Algorithm~\ref{alg:pcoco} is a Dini stationary point of $F(\bm\beta)$.
	\end{enumerate}
	\end{theorem}
\subsection{Connection to trimmed estimation}
In trimmed least squares (LS), the first step is to compute the residuals from a LS fit. Next, we identify and remove the outliers with large absolute residuals. Finally, we recalculate the LS solution using the remaining observations \citep{ruppert1980trimmed}. This estimator can be obtained using the IRCO with a concave tcave function and the initial estimator being the simple LS solution.

The CC-estimators are also closely related to least trimmed squares (LTS) estimator, which should not be confused with the trimmed LS. 
Instead of using all $n$ observations to calculate the          regression coefficients, a LTS estimator selects a subset of $\eta$             observations (where $\eta < n$) that result in the smallest sum of squared      residuals (least squares) among all possible combinations.
See \citet{maronna2019robust} and references therein. 

To illustrate the connection between Algorithm~\ref{alg:pcoco} and LTS, we will explicitly present Algorithm~\ref{alg:cocots} for the concave component tcave with $g(z) = \min(\sigma, z)$. This results in the IRCO for the truncation-stationary (IRCOTS) algorithm. In this case, we can obtain the total number of $z_i$ trimmed by $\sigma$ in Step 4:
\begin{equation*}
\eta^{(k+1)} = \#\{v_i^{(k+1)} = -1, i = 1, ..., n\}.
\end{equation*}
The data-driven value of $\eta^{(k+1)}$ is unspecified but can be computed using the fixed truncation parameter $\sigma$, which is why it is named truncation-stationary. Next, we modify the IRCOTS algorithm to make the estimator similar to the LTS estimator. Specifically, we adjust Step 4 in Algorithm~\ref{alg:cocots} such that $\eta^{(k+1)} = \eta$ for all $k$. This modification allows the location of truncation to change in each iteration.

By doing this, Algorithm~\ref{alg:cocotv} seeks a solution for the trimmed estimator as follows:
\begin{equation*}
\hat{\bm\beta} = \argmin_{\bm\beta} \sum_{i \in H} s(u_i(\bm\beta)) + \Lambda(\bm\beta),
\end{equation*}
where $H \subseteq \{1, ..., n\}$ and $|H| = \eta$. This equation represents the trimmed estimator. Finally, the IRCOTV (IRCO for truncation-varying) algorithm with $s(u) = u^2/2$ and LASSO penalty is the same as the algorithm for penalised LTS in \citet{alfons2013sparse}.
\begin{algorithm}[!htbp]
  \begin{algorithmic}[1]
    \caption{IRCOTS}\label{alg:cocots}
    \STATE \textbf{Initialise} $\bm\beta^{(0)}$ and set $k=0$
    \REPEAT
    \STATE Compute $u_i(\bm\beta^{(k)})$ in (\ref{eqn:ui}) and $z_i=s(u_i(\bm\beta^{(k)})), i=1, ..., n$
    \STATE Compute $v_i^{(k+1)}=-\textbf{1}(z_i \leq \sigma)$
      \STATE Compute $\bm\beta^{(k+1)} = \argmin_{\bm\beta} \sum_{i=1}^n s(u_i(\bm\beta))(-v_i^{(k+1)})+\Lambda(\bm\beta)$
    \STATE $k= k+1$
    \UNTIL convergence of $\bm\beta^{(k)}$
  \end{algorithmic}
\end{algorithm}
\begin{algorithm}[!htbp]
  \begin{algorithmic}[1]
    \caption{IRCOTV}\label{alg:cocotv}
    \STATE \textbf{Initialise} $\bm\beta^{(0)}$ and set $k=0$
    \REPEAT
      \STATE Compute $u_i(\bm\beta^{(k)})$ in (\ref{eqn:ui}) and $z_i=s(u_i(\bm\beta^{(k)})), i=1, ..., n$
      \STATE Compute $v_i^{(k+1)}=-\textbf{1}(z_i \leq z_\eta)$, where $z_1 \leq z_2 ... \leq z_n$ are ordered statistics, $\eta\leq n$
      \STATE Compute $\bm\beta^{(k+1)} = \argmin_{\bm\beta} \sum_{i=1}^n s(u_i(\bm\beta))(-v_i^{(k+1)})+\Lambda(\bm\beta)$
    \STATE $k= k+1$
    \UNTIL convergence of $\bm\beta^{(k)}$
  \end{algorithmic}
\end{algorithm}

\section{Applications of CC-estimators}
We conduct our comparisons using both simulated and real data. The response variables in our experiments include continuous, binary, and count data. We choose the robustness parameter $\sigma$ following the guidelines in Remark~\ref{rmk:sigma} for Algorithm~\ref{alg:pcoco}. For penalised estimation, the penalty parameter is determined using data-driven methods described below.

To evaluate the variable selection performance in simulated data, we compute sensitivity (sen) and specificity (spc). Sensitivity measures the proportion of correctly selected predictors among the truly effective predictors, while specificity measures the proportion of correctly non-selected predictors among the truly ineffective predictors. A good estimator should have both sensitivity and specificity close to 1, indicating accurate and precise variable selection.

For more detailed information about the applications and additional results, please refer to the Supplementary Information.
\subsection{Robust least squares in regression}

\indent Example 1 (nonpenalised): 
Let $\vec{y}=\matrix{x}\tran\bm\beta+\bm\epsilon$, where $\bm\beta=(1.5, 0.5, 1, 1.5, 1)^\intercal, \bm\epsilon$ is a $n$-dimensional vector with elements $\epsilon_i$ following a normal distribution with mean 0 and standard deviation 0.5, $i=1, ..., n, \vec x_i \sim \mathrm{N}_5(\vec{0}, \bm\Sigma)$ with $\Sigma_{ij}=0.5^{|i-j|}$ for $i, j=1, ..., 5$. Training and test data are randomly generated with sample size 100, where training data are used for model estimation, and test data are used to evaluate prediction accuracy. Test data are not contaminated, and contamination mechanisms in the training data follow \citet{alfons2013sparse}:

(1) No contamination

(2) Vertical outliers: 10\% of the error terms follow $\mathrm{N}(20, 0.5^2)$ instead of $\mathrm{N}(0, 0.5^2)$.

(3) Vertical outliers + leverage points: in addition to (2), the 10\% contaminated data also have predictor variables distributed as $\mathrm{N}(50, 1)$, different from the rest of predictor variables. 

Gaussian-induced CC-estimators without penalty are compared with least squares, biweight regression and LTS based on the root mean squared prediction error (RMSE). The average is reported in Table~\ref{tab:ex1314} for 100 Monte Carlo simulation runs. The oracle estimator is the true parameter, which provides the best prediction from the simulations. The CC-estimators are comparable with alternative methods for clean data and robust to outliers except for the hcave, i.e.,
the Huber estimator. It is well known that the Huber loss is robust to vertical outliers but not leverage points. 

\indent Example 2 (penalised): 
Let $\vec{y}=\matrix{x}\tran\bm\beta+\bm\epsilon$, where $\beta_1=\beta_7=1.5, \beta_2=0.5, \beta_4=\beta_{11}=1$ and $\beta_j=0$ otherwise for $j=1, ..., p, \bm\epsilon$ is a $n$-dimensional vector with elements $\bm\epsilon_i$ following a normal distribution with mean 0 and standard deviation 0.5, $i=1, ..., n, \vec x_i\sim \mathrm{N}_p(\vec{0}, \bm\Sigma)$ with $\Sigma_{ij}=0.5^{|i-j|}$ for $i, j=1, ..., p, p=50$. We generate random samples and simulation scheme as in Example 1. 
After training the model with the training data, a separate         portion of the data, called the tuning set, is used to fine-tune penalty            parameters. The best penalty parameters are chosen to be with the smallest loss     values on the tuning set.

Gaussian-induced penalised CC-estimators are computed and are compared with penalised linear regression, robust Huber regression \citep{yi2017semismooth} and sparseLTS \citep{alfons2013sparse}.
 The results are summarised in Table~\ref{tab:ex1112}. The penalised CC-estimators are comparable with penalised linear regressions for clean data, and outperform or are comparable with penalised linear regressions, Huber and LTS with outliers. As expected, the Huber loss (hcave) is robust to vertical outliers but not leverage points. The SCAD CC-estimators are better than their corresponding LASSO estimators. 
\subsection{Robust logistic regression}\label{sec:sec42}
\begin{figure}[!htbp]
  \begin{center}
    \includegraphics[scale=0.8]{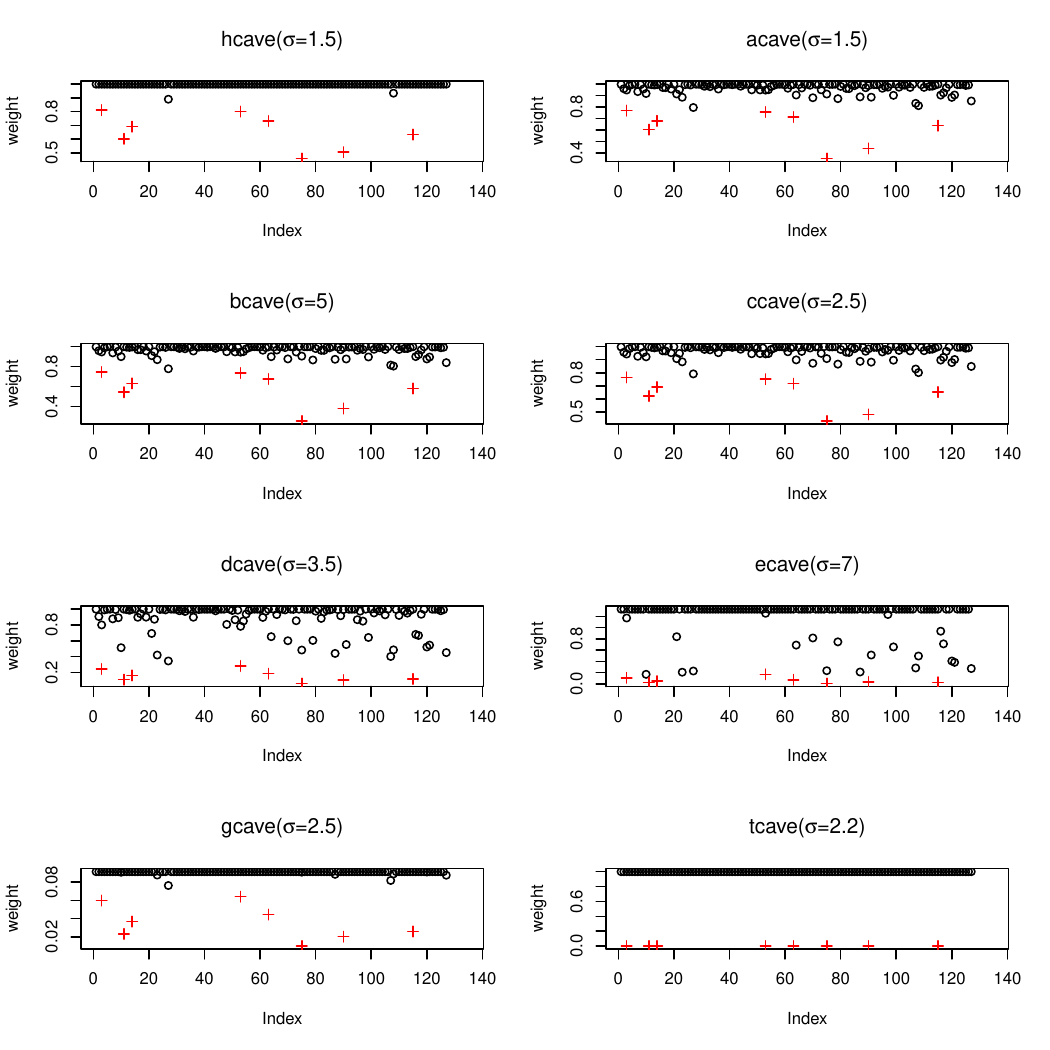}
    \caption{Robustness weights of logistic regression for the breastfeeding data.}
    \label{fig:brfeed}
  \end{center}
\end{figure}

In a survey conducted at a UK hospital, 135 expectant mothers were asked about their decision to breastfeed their babies or not. The survey also collected information on two-level predictive factors \citep{heritier2009robust}. We applied binomial-induced CC-estimators, which represent robust logistic regression, to the data and obtained robust weights. Figure~\ref{fig:brfeed} displays the robust weights for each individual. 
Notably, individuals 3, 11, 14, 53, 63, 75, 90, and 115 received the smallest weights in the robust logistic regression, which confirms the same results as \citet{heritier2009robust}, but our proposed CC-estimators achieve this using a simpler and more efficient approach.

Interestingly, some individuals showed counterintuitive results when using a logistic regression with large estimated probabilities (greater than or equal to 0.8) for either breastfeeding or not. Despite the high probabilities, these individuals made opposite decisions.

For variable selection, we developed a SCAD logistic regression, which offers sparser estimation than the LASSO estimator when the optimal penalty parameter $\lambda$ is determined using 10-fold cross-validation based on the maximum log-likelihood value. Using the optimal $\lambda$, we computed binomial-induced SCAD CC-estimators and obtained the estimated coefficients for the selected variables, as shown in Table~\ref{tab:brfeed}.

Comparing the coefficient of \texttt{smokenowYes} in the penalised logistic regression (which is $-2$), we found that the odds-ratio of a desire to breastfeed for a current smoking mother relative to a non-smoking mother is equal to $\exp(-2)=0.14$. However, the CC-estimators produced coefficients for \texttt{smokenowYes} that are less than $-2$, indicating that being a smoker during pregnancy has an even larger negative effect according to robust estimation.

Similarly, in all CC-estimators except for dcave, the odds-ratios of a desire to breastfeed for a non-White expecting mother relative to a White mother are larger than $\exp(1.94)=7$, which is derived from the penalised logistic regression.

These results highlight the benefits of using robust estimators, such as CC-estimators, in providing more accurate and reliable estimates in the presence of potential outliers and complex relationships in the data.
\subsection{Robust Poisson regression}\label{sec:sec43}
\begin{figure}[!htbp]
  \begin{center}
    \includegraphics[scale=0.8]{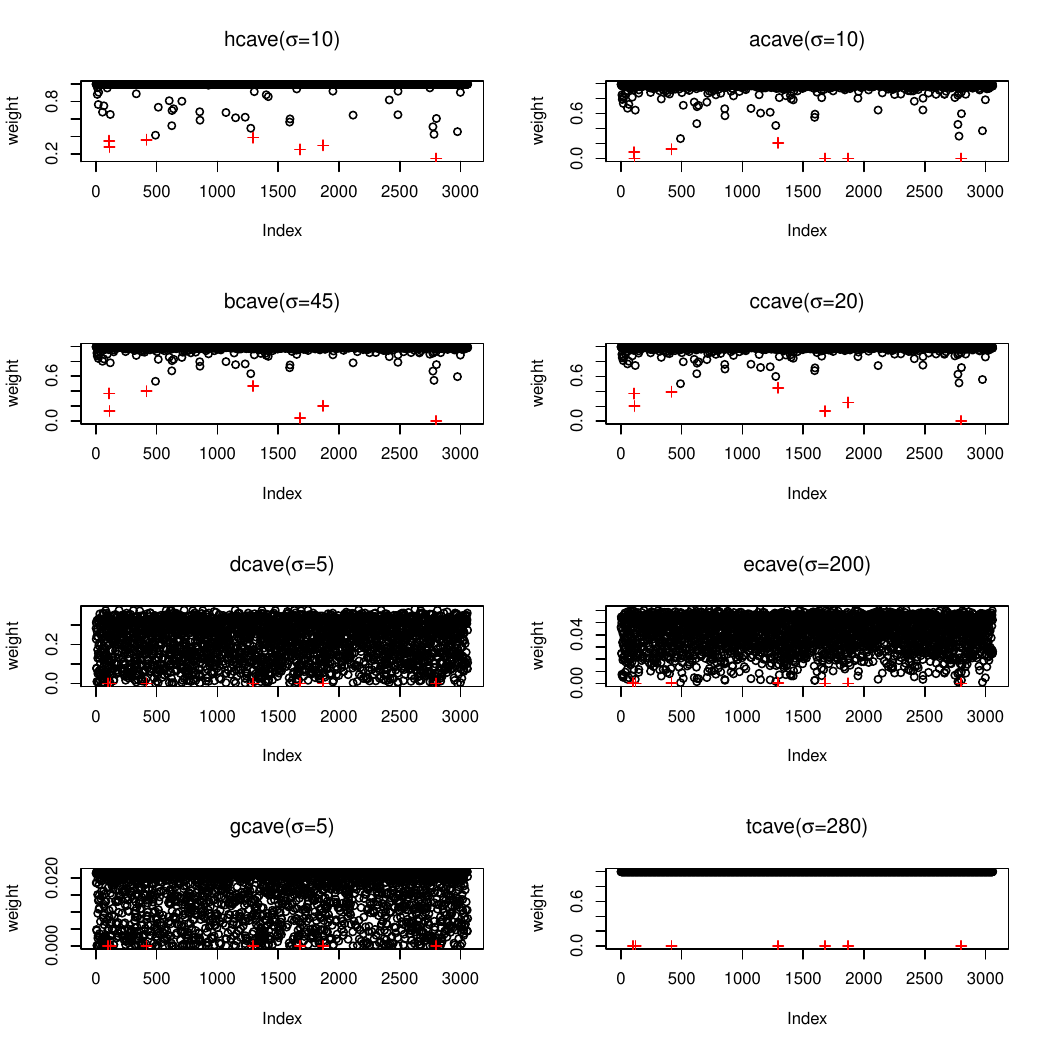}
    \caption{Robustness weights of Poisson regression for the doctor visits data.}
    \label{fig:docvisit}
  \end{center}
\end{figure}

In the study of health care utilisation among a cohort of 3066 Americans over the age of 50 \citep{heritier2009robust}, the outcome of interest was the number of doctor office visits. The survey also contained 24 predictors related to demographic, health needs, and economic access. We employed Poisson-induced CC-estimators, also known as robust Poisson regression, to analyse the data. Figure~\ref{fig:docvisit} displays the corresponding robust weights, and interestingly, we observed that the seven smallest weights correspond to subjects with 200, 208, 224, 260, 300, 365, and 750 doctor visits in two years, which aligns with the findings of \citet{heritier2009robust} using a more complex M-estimator.

To determine the optimal penalty parameter $\lambda$ for the ordinary SCAD Poisson regression, we conducted a 10-fold cross-validation, maximising the log-likelihood value. Utilising this selected $\lambda$ value, we computed Poisson-induced SCAD CC-estimators. The estimated coefficients of the selected variables are presented in Table~\ref{tab:docvisit}.

In both the penalised Poisson regression and our Poisson-induced CC-estimators, we observed a negative coefficient for the variable \texttt{age}, suggesting that older patients tend to consume fewer healthcare resources. This finding is consistent with the statistically significant coefficient of -0.005 reported by \citet{heritier2009robust} using their M-estimator. However, our approach provides a simpler estimation procedure without the need for a complex estimator.

\section{Discussion}
It is important to emphasise that the main objective of this article is to unify various robust loss           functions existing in the literature. Additionally, the article aims to extend the application of these loss           functions to penalised estimation for shrinkage parameter estimation and variable selection. The article also          provides a single computing algorithm that ensures a monotonically decreasing trend in the robust loss values.
 The IRCO algorithm, which is utilised in this work, holds a practical interpretation for outlier detection. The data-dependent weights employed in the algorithm are linked to outliers, where more extreme observations are assigned       smaller weights.
 
In regression models, when the random error terms have a symmetric distribution, the proposed estimators may   hold Fisher-consistency with random predictors \citep[Section 10.11]{maronna2019robust}.     In the context of GLMs,   this class of estimators can be seen as an extension of Pregibon's work from 1982. However, these estimators do not    exhibit Fisher-consistency when dealing with random predictors. See \citet[p. 277]{maronna2019robust} and the cited    references for further details on this aspect.
 Despite its limitations, the proposed approach offers valuable insights and applications in robust statistical          modelling.

This paper proposes a large family of loss functions, the CC-family, which is a composite of concave functions $g(\cdot)$ and convex functions $s(\cdot)$. 
When applying the CC-family to real applications, the choice of $g(\cdot)$ and $s(\cdot)$ becomes crucial. Selecting appropriate functions can significantly impact model performance. To address this, one may determine an optimal member from the large family of robust loss functions based on model predictive power in applications \citep{Hast:esl:2009}.

In Sections~\ref{sec:sec42} and \ref{sec:sec43}, we aimed to develop predictive models while identifying potential outliers, comparing the results to those in \citet{heritier2009robust}. However, it's important to note that the studies had a limitation: there was no dedicated test dataset to assess and determine optimal models. To overcome this limitation, one could consider splitting the available data into training and test datasets for model evaluation. However, caution should be exercised
when comparing the results to Tables~\ref{tab:brfeed} and ~\ref{tab:docvisit} and Figures~\ref{fig:brfeed} and \ref{fig:docvisit}, as the new models have different sample sizes and possibly different coefficients, model selection results, and outliers.

Although a predictive modelling approach is standard in many cases, we have chosen not to pursue it in this article. Instead, we focus on the development and evaluation of the CC-family and the IRCO algorithm.

We propose potential avenues for further research on CC-estimators. One direction is to explore the efficiency of CC-estimators compared to standard estimators. Specifically, we can investigate the efficiency gains achieved by CC-estimators with concave component and various convex components listed in Tables~\ref{tab:gs} and \ref{tab:dfun}. Efforts can be made to develop adaptive LASSO CC-estimators, where weighted penalties are prescribed based on the estimated coefficients from a preliminary or initial fit of the model \citep{Zou:2006}. The IRCO can be utilised to handle the optimisation problem in adaptive LASSO and examine the properties of the resulting estimators. Oracle properties, similar to those established for adaptive LASSO M-estimators \citep{smucler2017robust}, could be explored for certain members of the CC-family.

Another potential research direction is to consider estimating scale parameters of the exponential family within the CC-family. Robust scale estimators could be developed to address this aspect of the estimation problem \citep{hampel1986robust}. These robust scale estimators may prove useful in enhancing the robustness and accuracy of the overall estimation process.

Expanding the convex component of the CC-family opens up possibilities for applying CC-estimators and the IRCO to various statistical applications. For instance, the combination of CC-estimators and decision tree learning-based boosting, a popular toolkit in machine learning \citep{wang2021unified}, could lead to novel and effective approaches for handling complex data analysis problems.

In summary, these potential research directions offer exciting opportunities to further explore and extend the CC-family and its associated estimation framework, providing new insights and practical solutions for robust statistical and machine learning applications.
\section{Acknowledgment}
The author would like to thank two referees     for their constructive comments, which have significantly contributed to improving the quality of this paper. This work was partially supported by the National       Institute of Diabetes and Digestive and Kidney Diseases of the National        Institutes of Health under Award Number R21DK130006.

\putbib[wangres]
\end{bibunit}
\begin{table}
\small
\caption{Composite loss functions with $\sigma > 0$ unless otherwise specified.} 
	\begin{center}
\begin{tabular}{ lccc}
	\hline\hline
    Type & Loss function $g(s(u))$ & $g(z)$ & $s(u)$\\
\hline
    \textbf{Regression}  & & &\\
    \hspace{3mm}Huber  &
        $\begin{cases}
            \frac{u^2}{2} & \text { if } |u| \leq \sigma,\\ 
            \sigma|u|-\frac{\sigma^2}{2} &\text{ if } |u| > \sigma.
        \end{cases}$ &
        $\begin{cases}
            z & \text { if } z \leq \sigma^2/2,\\ 
            \sigma (2z)^{\frac{1}{2}}-\frac{\sigma^2}{2} &\text{ if } z > \sigma^2/2.
        \end{cases}$ 
        &$\frac{u^2}{2}$
        \\
    \hspace{3mm}Andrews &
    $\begin{cases}
        \makecell{\sigma(1-\cos(\frac{u}{\sigma}))} \\
        {\text{\quad \;\;  if } |u| \leq\sigma\pi},\\
         2\sigma \text{ if } |u| > \sigma\pi.\\
    \end{cases}$
         &
    $\begin{cases}
        \makecell{        {\sigma}(1-\cos(\frac{(2z)^{\frac{1}{2}}}{{\sigma}}))}\\ {\text{\quad \;\ if } z \leq \sigma^2\pi^2/2},\\
         2\sigma \text{ if } z > \sigma^2\pi^2/2.\\
    \end{cases}$
         &$\frac{u^2}{2}$\\
    \hspace{3mm}Biweight &$1-(1-\frac{u^2}{\sigma^2})^3 I(|u| \leq \sigma)$&
    {$1-(1-\frac{2z}{\sigma^2})^3 I(z \leq \sigma^2/2)$} & $\frac{u^2}{2}$\\
    \hspace{3mm}ClossR  &$1 - \exp(\frac{-u^2}{2\sigma^2})$ & $1-\exp(\frac{-z}{\sigma^2})$ &$\frac{u^2}{2}$\\
\hline
    \textbf{Classification}  && &\\
    \hspace{3mm}Closs &$1 - \exp(\frac{-(1-u)^2}{2\sigma^2})$
    & $1-\exp(\frac{-z}{\sigma^2})$ & $\frac{(1-u)^2}{2}$\\ 
    \hspace{3mm}Rhinge &$1-\exp(-\frac{\max(0, 1-u)}{2\sigma^2})$ & 
    $1-\exp(\frac{-z}{2\sigma^2})$ 
    & $\max(0, 1-u)$\\
     \hspace{3mm}Thinge & \makecell{$\min(1-\sigma, \max(0, 1-u))$, \\$\sigma \leq 0$} & $\min(1-\sigma, z)$ & $\max(0, 1-u)$\\
     \hspace{3mm}Tlogit & \makecell{$\min (1-\sigma, \log(1 + \exp(-u)))$,\\ $\sigma \leq 0$} & $\min(1-\sigma, z)$ & $\log(1 + \exp(-u))$\\
     \hspace{3mm}Texp & \makecell{$\min (1-\sigma, \exp(-u))$,\\ $\sigma \leq 0$} & $\min(1-\sigma, z)$ & $\exp(-u)$\\
    \hspace{3mm}Dlogit  & \makecell{$\log\left(1+\exp(-u)\right)$\\$-\log\left(1+\exp(-u-\sigma)\right)$} &
    $\log(\frac{1+z}{1+z\exp(-\sigma)})$ & $\exp(-u)$\\ 
    \hspace{3mm}Gloss &$\frac{1}{\left(1+\exp(au)\right)^\sigma}, \sigma \geq 1, a > 0$ & $(\frac{z}{1+z})^{\sigma}$ &$\exp(-au)$\\
    \hspace{3mm}Qloss &$1-\int_{\infty}^{\frac{u}{\sigma}}\frac{1}{\sqrt{2\pi}}\exp(\frac{-x^2}{2})dx$ & $1-\frac{1}{\sqrt{\pi}}\int_{0}^{\frac{z}{\sigma^2}} \frac{\exp(-t)}{\sqrt{t}}dt$& $\frac{u^2}{2}$\\
    \hline
\hline
\end{tabular}
	\end{center}
\label{tab:tab1} 
\end{table}

\normalsize
\begin{table}
\caption{Concave component with $\sigma > 0$.} 
	\begin{center}
\begin{tabular}{lcc}
	\hline\hline
    Concave& $g(z), z \ge 0$&Source\\
\hline
     hcave&$\begin{cases}
            z & \text { if } z \leq \sigma^2/2,\\ 
            \sigma (2z)^{\frac{1}{2}}-\frac{\sigma^2}{2} &\text{ if } z > \sigma^2/2.
     \end{cases}$ &Huber \\
    acave&\hspace{3mm}$\begin{cases}
         {\sigma^2}(1-\cos(\frac{(2z)^{\frac{1}{2}}}{{\sigma}})) & \text{ if } z \leq \sigma^2\pi^2/2, \\
         2\sigma^2 &\text{ if } z > \sigma^2\pi^2/2.\\
    \end{cases}$&Andrews 
    \\
    bcave&\hspace{3mm}$\frac{\sigma^2}{6}\left(1-(1-\frac{2z}{\sigma^2})^3 I(z \leq \sigma^2/2)\right)$  &Biweight\\
    ccave&\hspace{3mm}$\sigma^2\left(1-\exp(\frac{-z}{\sigma^2})\right)$ &Closs\\ 
     dcave&\hspace{3mm}$\frac{1}{1-\exp(-\sigma)}\log(\frac{1+z}{1+z\exp(-\sigma)})$&Dlogit \\
     ecave&\hspace{3mm}$\begin{cases}
     \frac{2\exp(-\frac{\delta}{\sigma})}{\sqrt{\pi\sigma\delta}}z &\text{ if }z\leq \delta,\\
     \erf(\sqrt{\frac{z}{\sigma}})-\erf(\sqrt{\frac{\delta}{\sigma}})+\frac{2\exp(-\frac{\delta}{\sigma})}{\sqrt{\pi\sigma\delta}}\delta &\text{ if } z>\delta.
     \end{cases}$&Qloss\\
    gcave&\hspace{3mm}
     $\begin{cases}
     \frac{\delta^{\sigma-1}}{(1+\delta)^{\sigma+1}}z &\text{ if } z \leq \delta,\\
     \frac{1}{\sigma}(\frac{z}{1+z})^{\sigma}-\frac{1}{\sigma}(\frac{\delta}{1+\delta})^{\sigma}+\frac{\delta^{\sigma}}{(1+\delta)^{\sigma+1}} &\text{ if }z > \delta.
     \end{cases}$&Gloss \\
     &where
     $\delta=
     \begin{cases}
     \to 0+ &\text{ if } 0 < \sigma < 1,\\ 
     \frac{\sigma-1}{2} &\text{ if }\sigma \geq 1.
     \end{cases}
     $&\\
    tcave&\hspace{3mm}$\min(\sigma, z), \sigma \geq 1 \text{ for classification; }\sigma > 0 \text{ otherwise }$&Truncation \\
\hline
\hline
\end{tabular}
	\end{center}
\label{tab:gs} 
\end{table}

\begin{table}
\caption{Convex component.}
	\begin{center}
\begin{tabular}{lc}
	\hline\hline
   Convex&$s(u)$\\
   \hline
     Gaussian&$\frac{u^2}{2}$\\
     GaussianC&$\frac{(1-u)^2}{2}$\\ 
     Binomial &$\log(1+\exp(-u))$\\
     Exponential family &$-\left(\frac{yu-b(u)}{a(\phi)}+c(y, \phi)\right)$\\
    Hinge &$\max(0, 1-u)$\\
    
     $\epsilon$-insensitive&
    $\begin{cases}
    0 &\text{ if } |u| \leq \epsilon, \\
    |u|-\epsilon &\text{ if } |u| > \epsilon.
    \end{cases}$
    \\
 
\hline
\hline
\end{tabular}
	\end{center}
\label{tab:dfun} 
\end{table}

\begin{table}
\caption{Subdifferential of negative concave component.}
	\begin{center}
\begin{tabular}{ lc}
	\hline\hline
    Concave & $\partial (-g(z))$\\
\hline
    hcave  &
        $\begin{cases}
            -1 & \text { if } z \leq \sigma^2/2,\\ 
            -\sigma (2z)^{-\frac{1}{2}} &\text{ if } z > \sigma^2/2.
        \end{cases}$ 
        \\
    acave &
    $\begin{cases}
         -\frac{\sigma\sin(\frac{\sqrt{2z}}{{\sigma}})}{\sqrt{2z}} &{\text{ if } 0 < z \leq \sigma^2\pi^2/2},\\
     -1 &\text{ if } z = 0,\\
         0 &\text{ if } z > \sigma^2\pi^2/2.\\
    \end{cases}$
         \\
    bcave &
    $-\frac{1}{\sigma^4}
    (2z-\sigma^2)^2 \textbf{1}(z \leq \sigma^2/2)$ \\
    ccave  & $ -\exp({-\frac{z}{{\sigma}^2}})
    $\\
    dcave & $-\frac{\exp(\sigma)}{(z+1)(z+\exp(\sigma))}$\\
    ecave &$
    \begin{cases}
     -\frac{2}{\sqrt{\pi\sigma\delta}}\exp(\frac{-\delta}{\sigma}) &\text{ if }z\leq \delta,\\
    -\frac{2}{\sqrt{\pi\sigma z }}\exp(\frac{-z}{\sigma}) &\text{ if } z>\delta.
     \end{cases}$\\
    gcave & 
    $\begin{cases}
    -\frac{\delta^{\sigma-1}}{(\delta+1)^{\sigma+1}} &\text{ if } z \leq \delta,\\
    -\frac{ z^{\sigma-1}}{(z+1)^{\sigma+1}} &\text{ if } z > \delta.
    \end{cases}$\\
    tcave & 
    $\begin{cases}
         \{-1\} &\text{ if } z < \sigma,\\
         \{0\} &\text{ if } z > \sigma,\\
         [-1, 0] &\text{ if } z = \sigma.
    \end{cases}$\\
    \hline
\hline
\end{tabular}
	\end{center}
\label{tab:sub} 
\end{table}

\begin{table}[!tbp]
\caption{RMSE in Example 1\label{tab:ex1314}.} 
\begin{center}
\begin{tabular}{lrrr}
\hline\hline
\multicolumn{1}{l}{Method$(\sigma)$}&\multicolumn{1}{c}{No conta-}&\multicolumn{1}{c}{Vertical}&\multicolumn{1}{c}{Vertical+}\\
& mination & outliers & Leverage
\tabularnewline
\hline
$LS$&$0.51$&$2.44$&$3.43$\tabularnewline
Biweight&$0.51$&$0.51$&$0.51$\tabularnewline
LTS&$0.52$&$0.52$&$0.52$\tabularnewline
hcave(1.3)&$0.51$&$0.55$&$3.45$\tabularnewline
acave(0.9)&$0.51$&$0.51$&$0.51$\tabularnewline
bacve(4.7)&$0.51$&$0.51$&$0.51$\tabularnewline
ccave(1.5)&$0.51$&$0.51$&$0.51$\tabularnewline
dcave(0.5)&$0.51$&$0.52$&$0.52$\tabularnewline
ecave(1.5)&$0.52$&$0.52$&$0.52$\tabularnewline
gcave(1.5)&$0.51$&$0.51$&$0.51$\tabularnewline
tcave(1.0)&$0.51$&$0.51$&$0.51$\tabularnewline
Oracle&$0.50$&$0.50$&$0.50$\tabularnewline
\hline
\end{tabular}\end{center}
\end{table}

\begin{table}[!tbp]
\caption{Estimation and prediction in Example 2.\label{tab:ex1112}} 
\begin{center}
\begin{tabular}{lccccccccccc}
\hline\hline
\multicolumn{1}{l}{\bfseries Method$(\sigma)$}&\multicolumn{3}{c}{\bfseries No contamination}&\multicolumn{1}{c}{\bfseries }&\multicolumn{3}{c}{\bfseries Vertical outliers}&\multicolumn{1}{c}{\bfseries }&\multicolumn{3}{c}{\bfseries Vertical+Leverage}\tabularnewline
\cline{2-4} \cline{6-8} \cline{10-12}
\multicolumn{1}{l}{}&\multicolumn{1}{c}{RMSE}&\multicolumn{1}{c}{Sen}&\multicolumn{1}{c}{Spc}&\multicolumn{1}{c}{}&\multicolumn{1}{c}{RMSE}&\multicolumn{1}{c}{Sen}&\multicolumn{1}{c}{Spc}&\multicolumn{1}{c}{}&\multicolumn{1}{c}{RMSE}&\multicolumn{1}{c}{Sen}&\multicolumn{1}{c}{Spc}\tabularnewline
\hline
LS LASSO&$0.54$&$1$&$0.76$&&$2.96$&$0.63$&$0.84$&&$1.73$&$0.98$&$0.50$\tabularnewline
LS SCAD&$0.51$&$1$&$0.95$&&$2.98$&$0.57$&$0.89$&&$1.84$&$0.89$&$0.75$\tabularnewline
Huber LASSO&$0.54$&$1$&$0.75$&&$0.57$&$1.00$&$0.76$&&$2.71$&$0.46$&$0.95$\tabularnewline
SparseLTS&$0.62$&$1$&$0.92$&&$0.58$&$1.00$&$0.90$&&$0.58$&$1.00$&$0.89$\tabularnewline
hcave(0.5)LASSO&$0.54$&$1$&$0.75$&&$0.58$&$1.00$&$0.75$&&$1.84$&$0.97$&$0.53$\tabularnewline
hcave(0.5)SCAD&$0.52$&$1$&$0.96$&&$0.53$&$1.00$&$0.96$&&$1.90$&$0.88$&$0.72$\tabularnewline
acave(0.9)LASSO&$0.54$&$1$&$0.76$&&$0.55$&$1.00$&$0.77$&&$0.55$&$1.00$&$0.77$\tabularnewline
acave(0.9)SCAD&$0.51$&$1$&$0.95$&&$0.52$&$1.00$&$0.96$&&$0.51$&$1.00$&$0.96$\tabularnewline
bcave(4.7)LASSO&$0.54$&$1$&$0.76$&&$0.55$&$1.00$&$0.77$&&$0.55$&$1.00$&$0.77$\tabularnewline
bcave(4.7)SCAD&$0.51$&$1$&$0.96$&&$0.51$&$1.00$&$0.96$&&$0.52$&$1.00$&$0.95$\tabularnewline
ccave(1.5)LASSO&$0.54$&$1$&$0.75$&&$0.55$&$1.00$&$0.77$&&$0.55$&$1.00$&$0.77$\tabularnewline
ccave(1.5)SCAD&$0.51$&$1$&$0.95$&&$0.51$&$1.00$&$0.96$&&$0.51$&$1.00$&$0.96$\tabularnewline
dcave(0.5)LASSO&$0.54$&$1$&$0.76$&&$0.55$&$1.00$&$0.76$&&$0.55$&$1.00$&$0.79$\tabularnewline
dcave(0.5)SCAD&$0.51$&$1$&$0.96$&&$0.52$&$1.00$&$0.95$&&$0.53$&$1.00$&$0.95$\tabularnewline
ecave(9.0)LASSO&$0.54$&$1$&$0.74$&&$0.55$&$1.00$&$0.76$&&$0.54$&$1.00$&$0.82$\tabularnewline
ecave(9.0)SCAD&$0.52$&$1$&$0.95$&&$0.52$&$1.00$&$0.95$&&$0.52$&$1.00$&$0.95$\tabularnewline
gcave(1.5)LASSO&$0.54$&$1$&$0.75$&&$0.55$&$1.00$&$0.77$&&$0.54$&$1.00$&$0.80$\tabularnewline
gcave(1.5)SCAD&$0.51$&$1$&$0.96$&&$0.51$&$1.00$&$0.96$&&$0.54$&$1.00$&$0.95$\tabularnewline
tcave(2.5)LASSO&$0.54$&$1$&$0.76$&&$0.55$&$1.00$&$0.77$&&$0.54$&$1.00$&$0.80$\tabularnewline
tcave(2.5)SCAD&$0.51$&$1$&$0.95$&&$0.51$&$1.00$&$0.95$&&$0.51$&$1.00$&$0.96$\tabularnewline
Oracle&$0.50$&$1$&$1.00$&&$0.50$&$1.00$&$1.00$&&$0.50$&$1.00$&$1.00$\tabularnewline
\hline
\end{tabular}\end{center}
\end{table}

\begin{table}[!tbp]
\small
\caption{Estimates of robust penalised logistic regression for the breastfeeding data.\label{tab:brfeed}} 
\begin{center}
\begin{tabular}{lrrrrrrrrr}
\hline\hline
\multicolumn{1}{l}{Variable}&\multicolumn{1}{c}{logis}&\multicolumn{1}{c}{hcave}&\multicolumn{1}{c}{acave}&\multicolumn{1}{c}{bcave}&\multicolumn{1}{c}{ccave}&\multicolumn{1}{c}{dcave}&\multicolumn{1}{c}{ecave}&\multicolumn{1}{c}{gcave}&\multicolumn{1}{c}{tcave}\tabularnewline
\hline
(Intercept)&$ 0.10$&$-0.20$&$ 0.32$&$ 0.33$&$ 0.35$&$ 2.71$&$ 3.27$&$-0.70$&$-2.27$\tabularnewline
pregnancyBeginning&$ $&$ $&$ $&$ $&$ $&$ $&$ $&$ $&$ $\tabularnewline
howfedBreast&$ $&$ $&$ $&$ $&$ $&$ 0.12$&$ $&$ $&$ $\tabularnewline
howfedfrBreast&$ 1.05$&$ 1.42$&$ 1.19$&$ 1.21$&$ 1.18$&$ 0.03$&$ 0.05$&$ 1.76$&$ 1.27$\tabularnewline
partnerPartner&$ 0.48$&$ 0.24$&$ 0.20$&$ 0.13$&$ 0.22$&$ $&$ $&$ $&$ $\tabularnewline
smokenowYes&$-2.00$&$-2.31$&$-2.38$&$-2.44$&$-2.38$&$-3.89$&$-4.25$&$-2.69$&$-2.48$\tabularnewline
smokebfYes&$ $&$ $&$ $&$ $&$ $&$ $&$ $&$ $&$ $\tabularnewline
age&$ $&$ $&$ $&$ $&$ $&$ $&$ $&$ $&$ $\tabularnewline
educat&$ $&$ 0.03$&$ 0.01$&$ 0.01$&$ 0.01$&$ $&$ $&$ 0.06$&$ 0.16$\tabularnewline
ethnicNon-white&$ 1.94$&$ 2.49$&$ 2.52$&$ 2.64$&$ 2.48$&$ 1.16$&$ 2.45$&$ 3.25$&$ 3.59$\tabularnewline
\hline
\end{tabular}\end{center}
\end{table}

\normalsize
\begin{table}[!tbp]
\caption{Estimates of robust penalised Poisson regression for the doctor visits data.\label{tab:docvisit}} 
\begin{center}
\begin{tabular}{lrrrrrrrrr}
\hline\hline
\multicolumn{1}{l}{Variable}&\multicolumn{1}{c}{Poisson}&\multicolumn{1}{c}{hcave}&\multicolumn{1}{c}{acave}&\multicolumn{1}{c}{bcave}&\multicolumn{1}{c}{ccave}&\multicolumn{1}{c}{dcave}&\multicolumn{1}{c}{ecave}&\multicolumn{1}{c}{gcave}&\multicolumn{1}{c}{tcave}\tabularnewline
\hline
(Intercept)&$1.86$&$1.99$&$1.98$&$1.98$&$1.98$&$1.83$&$1.88$&$1.78$&$1.97$\tabularnewline
age&$-4\times$&$$&$$&$-5\times$&$-4\times$&$$&$$&$$&$$\tabularnewline
&$ 10^{-3}$&$$&$$&$10^{-5}$&$10^{-5}$&$$&$$&$$&$$\tabularnewline
gender&$$&$$&$$&$$&$$&$$&$$&$$&$$\tabularnewline
race&$$&$$&$$&$$&$$&$$&$$&$$&$$\tabularnewline
hispan&$$&$$&$$&$$&$$&$$&$$&$$&$$\tabularnewline
marital&$$&$$&$$&$$&$$&$$&$$&$$&$$\tabularnewline
arthri&$0.03$&$0.04$&$0.05$&$0.04$&$0.03$&$0.03$&$0.03$&$0.03$&$0.06$\tabularnewline
cancer&$0.07$&$0.03$&$0.03$&$0.02$&$0.02$&$$&$0.01$&$$&$0.03$\tabularnewline
hipress&$0.12$&$0.11$&$0.08$&$0.12$&$0.13$&$0.05$&$0.07$&$0.07$&$0.08$\tabularnewline
diabet&$0.30$&$0.22$&$0.20$&$0.20$&$0.19$&$0.03$&$0.07$&$0.01$&$0.24$\tabularnewline
lung&$$&$0.01$&$0.03$&$0.03$&$0.02$&$$&$$&$$&$0.03$\tabularnewline
heart&$0.29$&$0.32$&$0.33$&$0.33$&$0.33$&$0.36$&$0.35$&$0.34$&$0.33$\tabularnewline
stroke&$$&$0.05$&$0.07$&$0.07$&$0.06$&$$&$$&$$&$0.13$\tabularnewline
psych&$0.25$&$0.27$&$0.28$&$0.29$&$0.28$&$0.03$&$0.08$&$0.02$&$0.31$\tabularnewline
iadla1&$$&$$&$$&$$&$$&$$&$$&$$&$$\tabularnewline
iadla2&$$&$$&$$&$$&$$&$$&$$&$$&$$\tabularnewline
iadla3&$$&$$&$$&$$&$$&$$&$$&$$&$$\tabularnewline
adlwa1&$0.37$&$0.25$&$0.14$&$0.27$&$0.27$&$$&$0.05$&$$&$0.20$\tabularnewline
adlwa2&$0.68$&$0.44$&$0.37$&$0.39$&$0.40$&$$&$0.36$&$$&$0.37$\tabularnewline
adlwa3&$0.64$&$0.54$&$0.49$&$0.51$&$0.52$&$0.60$&$0.59$&$0.65$&$0.46$\tabularnewline
edyears&$$&$$&$$&$$&$$&$$&$$&$$&$$\tabularnewline
feduc&$$&$$&$$&$$&$$&$$&$$&$$&$$\tabularnewline
meduc&$$&$$&$$&$$&$$&$$&$$&$$&$$\tabularnewline
log(income + 1)&$0.04$&$$&$$&$$&$$&$$&$$&$$&$$\tabularnewline
insur&$0.02$&$$&$$&$$&$$&$$&$$&$$&$$\tabularnewline
\hline
\end{tabular}\end{center}
\end{table}

\newpage

\clearpage
 \setcounter{page}{0}
     \pagenumbering{arabic}
     \setcounter{page}{1}
 \begin{center}
     {\LARGE Unified Robust Estimation}\\
 Supplementary Information\\
 Zhu Wang\\
 The University of Tennessee Health Science Center\\
 E-mail: zwang145@uthsc.edu
 \end{center}

\begin{appendices}
\begin{bibunit}[apalike]
\section{Comments and additional applications}
    \subsection{Comments to Section 4}
             In the simulation study, all CC-estimators produce almost           identical        results except for hcave when both response and predictor          variables have outliers.
         However, in
    the real example, especially for the doctor visits data, the estimated           coefficients and the robustness weights are sometimes largely different between     these CC-estimators     (e.g. tcave
         and gcave). There are at least two           reasons that could contribute  to the differences.
             
             First, penalty parameters are selected differently. In Example 2,   tuning data are utilised to help select the best penalty parameters with the        smallest robust loss values. However, for the real data analysis, such as doctor    visits data, the same penalty parameter is utilised, obtained from an ordinary      SCAD Poisson regression using a 10-fold cross-validation. This approach explicitly  compares the robust loss functions and the traditional loss function when the       penalty and its
 associated parameter are the same. The choice of the method may depend on   the specific purposes of the analysis.
 
             Second, it is expected that the analysis from different methods     can generate different coefficients and weights. As shown in Figure 2, which is     derived from Table 4, 
         tcave can only provide weights of 0 or 1,     unless in a degenerative      case where $z = \sigma$, which has a        probability of 0 to occur since $z$ is  continuous, while other concave functions can provide values in the whole range     of       $[0, 1]$.

    \subsection{Robust least squares in classification}

Example 3: Predictor variables $(x_1, x_2)$ are uniformly sampled      from a unit disk $x_1^2+x_2^2 \leq 1$ and $y=1$ if $x_1 \geq x_2$ and -1        otherwise.
We also generate 18 noise variables from uniform[-1, 1]. To add outliers, we randomly select \texttt{v} percent of the data and switch their class labels. The training/tuning/test sample sizes are $n=100/100/10,000$. 

We evaluate GaussianC-induced CC-estimators, i.e., the Gaussian-induced composite loss with $y\in\{+1, -1\}$. No-intercept models  are adopted for more accurate prediction. The penalised least squares method is also employed along with the optimal Bayes classifier. 
The results are demonstrated in Table~\ref{tab:ex1}. It is clear that the CC-estimators are better resistant to outliers than the LS estimators, and the SCAD estimators are better than the LASSO counterparts.  
\begin{table}[!tbp]
\small
\caption{Mean test errors, sensitivity and specificity in Example 3.\label{tab:ex1}} 
\begin{center}
\begin{tabular}{lrrrcrrrcrrr}
\hline\hline
\multicolumn{1}{l}{\bfseries Method$(\sigma)$}&\multicolumn{3}{c}{\bfseries No contamination}&\multicolumn{1}{c}{\bfseries }&\multicolumn{3}{c}{\bfseries 10\% contamination}&\multicolumn{1}{c}{\bfseries }&\multicolumn{3}{c}{\bfseries 20\% contamination}\tabularnewline
\cline{2-4} \cline{6-8} \cline{10-12}
\multicolumn{1}{l}{}&\multicolumn{1}{c}{Error}&\multicolumn{1}{c}{Sen}&\multicolumn{1}{c}{Spc}&\multicolumn{1}{c}{}&\multicolumn{1}{c}{Error}&\multicolumn{1}{c}{Sen}&\multicolumn{1}{c}{Spc}&\multicolumn{1}{c}{}&\multicolumn{1}{c}{Error}&\multicolumn{1}{c}{Sen}&\multicolumn{1}{c}{Spc}\tabularnewline
\hline
LS LASSO&$0.023$&$1$&$0.94$&&$0.137$&$1$&$0.87$&&$0.252$&$1$&$0.86$\tabularnewline
LS SCAD&$0.010$&$1$&$0.96$&&$0.131$&$1$&$0.90$&&$0.251$&$1$&$0.85$\tabularnewline
hcave(1)LASSO&$0.027$&$1$&$0.97$&&$0.135$&$1$&$0.90$&&$0.248$&$1$&$0.84$\tabularnewline
hcave(1)SCAD&$0.017$&$1$&$0.99$&&$0.120$&$1$&$0.99$&&$0.224$&$1$&$0.97$\tabularnewline
acave(1)LASSO&$0.029$&$1$&$0.98$&&$0.137$&$1$&$0.90$&&$0.251$&$1$&$0.84$\tabularnewline
acave(1)SCAD&$0.018$&$1$&$0.99$&&$0.121$&$1$&$0.98$&&$0.227$&$1$&$0.97$\tabularnewline
bcave(3.5)LASSO&$0.029$&$1$&$0.98$&&$0.137$&$1$&$0.90$&&$0.251$&$1$&$0.84$\tabularnewline
bcave(3.5)SCAD&$0.018$&$1$&$0.99$&&$0.121$&$1$&$0.99$&&$0.227$&$1$&$0.97$\tabularnewline
ccave(1.5)LASSO&$0.030$&$1$&$0.98$&&$0.137$&$1$&$0.90$&&$0.250$&$1$&$0.84$\tabularnewline
ccave(1.5)SCAD&$0.020$&$1$&$0.99$&&$0.121$&$1$&$0.99$&&$0.227$&$1$&$0.96$\tabularnewline
dcave(4.5)LASSO&$0.032$&$1$&$0.98$&&$0.137$&$1$&$0.91$&&$0.249$&$1$&$0.84$\tabularnewline
dcave(4.5)SCAD&$0.020$&$1$&$0.99$&&$0.122$&$1$&$0.99$&&$0.229$&$1$&$0.95$\tabularnewline
ecave(9)LASSO&$0.029$&$1$&$0.96$&&$0.136$&$1$&$0.91$&&$0.248$&$1$&$0.87$\tabularnewline
ecave(9)SCAD&$0.017$&$1$&$0.99$&&$0.120$&$1$&$0.98$&&$0.226$&$1$&$0.95$\tabularnewline
gcave(1.5)LASSO&$0.029$&$1$&$0.96$&&$0.135$&$1$&$0.90$&&$0.246$&$1$&$0.84$\tabularnewline
gcave(1.5)SCAD&$0.018$&$1$&$0.99$&&$0.120$&$1$&$0.99$&&$0.226$&$1$&$0.96$\tabularnewline
tcave(1)LASSO&$0.027$&$1$&$0.97$&&$0.129$&$1$&$0.91$&&$0.240$&$1$&$0.84$\tabularnewline
tcave(1)SCAD&$0.017$&$1$&$0.99$&&$0.117$&$1$&$0.97$&&$0.222$&$1$&$0.95$\tabularnewline
Bayes&$0.000$&$1$&$1.00$&&$0.100$&$1$&$1.00$&&$0.200$&$1$&$1.00$\tabularnewline
\hline
\end{tabular}\end{center}
\end{table}

\normalsize
\subsection{Robust SVM}\label{sec:svm}
A dataset
concerns Australian credit card applications for 690 samples with a good mix of 14 predictors -- continuous, nominal with small numbers of values, and nominal with larger numbers of values \citep{lichman2013uci}.
The hinge-induced CC-estimators, i.e., robust SVM, are utilised to predict credit card approval. We use 10-fold cross validation for model training and evaluation. We randomly choose 70\% of a fold with $n=690\times 0.9\times 0.7$ as training data, the remaining 30\% of a fold as tuning data with $n=690\times 0.9\times 0.3$ for hyper-parameters determinations. The test errors are then computed from the test data with $n=690\times 0.1$. This process is repeated 10 times based on the cross-validation scheme. To study robustness of algorithms, 15\% of credit card approval decision is randomly flipped in the training and tuning data.
We adopt the nonlinear Gaussian kernel in the SVM. From Table~\ref{tab:svmex1}, the CC-estimators are comparable to the SVM with clean data, and more accurate with contaminated data. For data with outliers, the averages number of support vectors from the CC-estimators are smaller than the SVM. That is, many more observations in the standard SVM are involved in determining the classification rule, which is not preferred. 
\begin{table}[!tbp]
\caption{Average test error rate and support vectors for credit card applications with different percentage of contamination (conta). \label{tab:svmex1}} 
\begin{center}
\begin{tabular}{lrrcrr}
\hline\hline
\multicolumn{1}{l}{\bfseries Method$(\sigma)$}&\multicolumn{2}{c}{\bfseries No conta}&\multicolumn{1}{c}{\bfseries }&\multicolumn{2}{c}{\bfseries 15\% conta}\tabularnewline
\cline{2-3} \cline{5-6}
\multicolumn{1}{l}{}&\multicolumn{1}{c}{Error}&\multicolumn{1}{c}{\#SV}&\multicolumn{1}{c}{}&\multicolumn{1}{c}{Error}&\multicolumn{1}{c}{\#SV}\tabularnewline
\hline
SVM&$0.144$&$274$&&$0.165$&$366$\tabularnewline
hcave(0.8)&$0.142$&$256$&&$0.148$&$306$\tabularnewline
acave(0.8)&$0.148$&$241$&&$0.158$&$311$\tabularnewline
bcave(4.8)&$0.145$&$275$&&$0.152$&$340$\tabularnewline
ccave(2.2)&$0.138$&$278$&&$0.152$&$338$\tabularnewline
dcave(2.6)&$0.138$&$244$&&$0.146$&$303$\tabularnewline
ecave(6.8)&$0.139$&$227$&&$0.145$&$294$\tabularnewline
gcave(1)&$0.149$&$211$&&$0.148$&$300$\tabularnewline
tcave(1.4)&$0.138$&$242$&&$0.154$&$244$\tabularnewline
\hline
\end{tabular}\end{center}
\end{table}

\subsection{Robust SVM regression}
The Boston housing data include 506 housing values and 14 predictors in suburbs of Boston \citep{lichman2013uci}.
We compute $\epsilon$-insensitive-induced CC-estimators, i.e., robust SVM regression, to predict the housing prices. We use 10-fold cross validation as in the previous example. To study robustness of algorithms, 10\% of housing values are randomly multiplied by 10 in the training and tuning data. The optimal hyper-parameters of the Gaussian kernel minimise the RMSE in the tuning data without outliers. In the contaminated data, these parameters are based on 90\% trimmed RMSE. The results are summarised in
Table~\ref{tab:svmex2}. The RMSEs are comparable in clean data while the CC-estimators are much robust than the SVM regression with contaminated data. The number of SVs are similar in the clean data, while seven out of eight CC-estimators have smaller SVs with contaminated data.
\begin{table}[!htbp]
\caption{Average RMSE and \# support vectors for Boston housing prices with different percentage of contamination (conta). \label{tab:svmex2}} 
\begin{center}
\begin{tabular}{lrrcrr}
\hline\hline
\multicolumn{1}{l}{\bfseries Method$(\sigma)$}&\multicolumn{2}{c}{\bfseries No conta}&\multicolumn{1}{c}{\bfseries }&\multicolumn{2}{c}{\bfseries 10\% conta}\tabularnewline
\cline{2-3} \cline{5-6}
\multicolumn{1}{l}{}&\multicolumn{1}{c}{RMSE}&\multicolumn{1}{c}{\#SV}&\multicolumn{1}{c}{}&\multicolumn{1}{c}{RMSE}&\multicolumn{1}{c}{\#SV}\tabularnewline
\hline
SVM&$3.60$&$190$&&$4.60$&$120$\tabularnewline
hcave(5)&$3.60$&$190$&&$4.40$&$97$\tabularnewline
acave(10)&$3.60$&$190$&&$4.50$&$100$\tabularnewline
bcave(24)&$3.60$&$190$&&$4.20$&$100$\tabularnewline
ccave(8)&$3.60$&$190$&&$4.30$&$ 91$\tabularnewline
dcave(10)&$3.70$&$180$&&$4.10$&$ 88$\tabularnewline
ecave(5)&$3.70$&$190$&&$4.30$&$ 87$\tabularnewline
gcave(20)&$3.70$&$180$&&$4.20$&$ 85$\tabularnewline
tcave(200)&$3.60$&$190$&&$4.20$&$140$\tabularnewline
\hline
\end{tabular}\end{center}
\end{table}

\clearpage
\section{Some theoretical background}\label{app:back}
\subsection{Regression M-estimators}
Consider nonpenalised robust linear regression with twice differentiable functions $g$ and $s$. A solution to $\argmin F(\bm\beta)$ can be obtained from the estimation equation:
\begin{equation*}
    \sum_{i=1}^n \Gamma'(r_i(\bm\beta))\vec{x_i} =0,
\end{equation*}
where $r_i(\bm\beta)=y_i-\vec x_i\tran \bm\beta$. 
While statistical inference is beyond the scope of the current paper, a brief summary may provide relevant insights. A different M-estimator based on the MLE can be derived \citep[Section 4.4]{maronna2019robust}. 
Suppose that
    $y_i=\vec x_i\tran\bm\beta+\epsilon_i$, $\vec x=(\vec x_1, ..., \vec x_n)\tran$ is fixed, $\epsilon_i$ has a probability density $\frac{1}{\phi}f(\frac{\mu}{\phi})$ for known scale $\phi$ such that $\Gamma=-\log f$, 
    $E(\Gamma'(\mu/\phi))=0$, and mild regularity conditions hold on the design matrix $\vec x$. If $\bm\beta^\ast$ satisfies the estimation equation
\begin{equation*}
    \sum_{i=1}^n \Gamma'\left(\frac{r_i(\bm\beta^\ast)}{\phi}\right)\vec{x_i} =0,
\end{equation*}
    then $\bm\beta^\ast$ is consistent for $\bm\beta$ and has the asymptotic normal distribution given by
    \begin{equation*}
        \bm\beta^\ast \xrightarrow{d} \mathcal{N}(\beta,v(\vec x\tran\vec x)^{-1}),
    \end{equation*}
        where
        \begin{equation*}
            v=\phi^2\frac{E(\Gamma'(\mu/\phi)^2)}{\left(E\Gamma''(\mu/\phi)\right)^2}.
        \end{equation*}
See \citet[Section 4.4.1]{maronna2019robust}.
\subsection{Dini stationary point}
\citet{clarke2013functional} discussed generalised derivatives for  nonsmooth nonconvex         functions. Consider $f: \mathbb{R}^m \to \mathbb{R}$.
             The lower directional Dini derivative of $f$ at $x$ in the              direction        $\varepsilon$ is defined below:
  \begin{equation*}
  f'_D(x; \varepsilon)\triangleq\liminf\limits_{\tau\to 0+}\frac{f(x+\tau            \varepsilon)-f(x)}{\tau}.
  \end{equation*}
 The point $x$ is a Dini stationary point of $f(\cdot)$ if $f'_D(x;                  \varepsilon) \geq 0, \varepsilon \in \mathbb{R}^m$.

\section{Proofs}\label{app:pro}

\noindent{\textbf{Proof of Theorem~\ref{thm:linear}}}

We only need to show that $g$ satisfies requirement (i) in Definition~\ref{assu:cc}. Suppose $z_1 < z_2$ for $z_1, z_2 \in \text{dom }g$, we then have $g_1(z_1) \leq g_1(z_2), g_2(z_1) \leq g_2(z_2)$ since $g_1$ and $g_2$ satisfy requirement (i) in Definition~\ref{assu:cc}. Hence $c_1g_1(z_1)+c_2g_2(z_1) \leq c_1g_1(z_2)+c_2g_2(z_2)$, or $g$ is nondecreasing. Following \citet[Lemma 3.1.9]{nesterov2004introductory}, $-g$ is closed convex and (\ref{eqn:linear}) holds. 
\qed

\noindent{\textbf{Proof of Theorem~\ref{thm:max}}}

    It is simple algebra to show that $g$ is nondecreasing. 
    Since $g=\min_{1 \leq i \leq m} g_i$, we get $-g=\max_{1\leq i \leq m}(-g_i)$.
    Following \citet[Lemma 3.1.10]{nesterov2004introductory}, $-g$
    is closed convex and (\ref{eqn:max}) holds.
\qed

\noindent{\textbf{Proof of Theorem~\ref{thm:concave11}}}

By assumption we have a well-defined function composition 
\begin{equation*}\label{eqn:lemconc_11}
  \Gamma(u)=g(s(u)).
\end{equation*}
It is simple algebra to show
\begin{equation}\label{eqn:lemconc_41}
  \Gamma''(u)=g''(s(u))(s'(u))^2+\frac{s''(u)}{s'(u)}\Gamma'(u).
\end{equation}
Suppose 
\begin{equation}\label{eqn:lemconc_51}
  \Gamma''(u) \leq \frac{s''(u)}{s'(u)}\Gamma'(u).
\end{equation}
From (\ref{eqn:lemconc_41}) we must have
\begin{equation*}
    g''(s(u))(s'(u))^2 \leq 0.
\end{equation*}
Since $s'(u) \neq 0$ by assumption, $g''(s(u)) \leq 0 \ \text{for every } u$ holds, or $g$ is concave. Conversely, if $g$ is concave, $g''(s(u)) \leq 0$ for every $u$, thus (\ref{eqn:lemconc_51}) holds. 
\qed

\noindent{\textbf{Proof of Theorem~\ref{thm:concave2}}}

We apply similar arguments as in \citet[page 35]{hiriart1993convex}. Suppose 
\begin{equation}\label{eqn:tmp1}
    \frac{s''(u)}{s'(u)}\Gamma'(u) \geq \Gamma''(u)
\end{equation}
holds piecewisely. Following the proof of Theorem~\ref{thm:concave11}, $g''(s(u)) \leq 0$ holds piecewisely. Since $g$ has decreasing slopes, then $g$ is concave. Conversely, if $g$ is concave, $g''(s(u)) \leq 0$ holds piecewisely. Hence (\ref{eqn:tmp1}) is valid as in the proof of Theorem~\ref{thm:concave11}.
\qed

\noindent{\textbf{Proof of Theorem~\ref{thm:fisher}}}

\begin{enumerate}[(i)]
  \item From condition~\ref{con1}, we know that $s(u) < s(-u), \ u > 0$. Thus $\Gamma(u)=g(s(u)) < g(s(-u))=\Gamma(-u), \ \text{for every } u > 0$ since $g$ is increasing from condition~\ref{con3}. 
        Furthermore, $\Gamma'(0)=g'(s(0))s'(0)\neq 0$ exists from conditions~\ref{con2} and \ref{con4}. 
  We conclude that $\Gamma=g\circ s$ satisfies the assumptions of Theorem 3.1 in \citet{lin2004note}, thus $\Gamma$ is Fisher-consistent.
  
\item Note that $E(\Gamma(Yf(X)))=E(E(\Gamma(Yf(X)|X=x)))$, we can minimise $E(\Gamma(Yf(X)))$ by minimising $E(\Gamma(Yf(X))|X=x)$ for every $x$. For any fixed $x$, $E(\Gamma(Yf(X))|X=x)=p(x)\Gamma(f(x))+(1-p(x))\Gamma(-f(x))$. We search $w^\ast=\argmin_w V(w)$, where
  \begin{equation*}
    V(w)=p(x)\Gamma(w)+(1-p(x))\Gamma(- w).
  \end{equation*}
  We have 
  \begin{equation*}
    V(-w)=p(x)\Gamma(-w)+(1-p(x))\Gamma(w).
  \end{equation*}
  The last two equations lead to
  \begin{equation*}\label{eqn:thm1.10}
    V(w)-V(-w)=(2p(x)-1)(\Gamma(w)-\Gamma(-w)).
  \end{equation*}
  From the definition of $w^\ast$, we obtain
    \begin{equation*}\label{eqn:thm1.12}
    V(w^\ast)-V(-w^\ast)=(2p(x)-1)(\Gamma(w^\ast)-\Gamma(-w^\ast)) \leq 0.
  \end{equation*}
        If $p(x) > \frac{1}{2}$, we have
    \begin{equation*}\label{eqn:thm1.14}
    \Gamma(w^\ast)-\Gamma(-w^\ast) \leq 0.
  \end{equation*}
        Since $\Gamma$ is non-increasing from condition~\ref{con6}, we have
    \begin{equation*}\label{eqn:thm1.16}
    w^\ast \geq -w^\ast,
  \end{equation*}
        which implies $w^\ast \geq 0$. Similarly, we get $w^\ast \leq 0$ if $p(x) < \frac{1}{2}$. 
        Hence, it is sufficient to show that $w=0$ is not a minimiser of $V(w)$. 
        In the following, we consider two cases. 
       If $\sigma=1$, from condition~\ref{con5}, we obtain
        \begin{equation*}
    \begin{aligned}
      V(0)&=p(x)g(s(0))+(1-p(x))g(s(0))\\
      &    >p(x)g(s(1))+(1-p(x))g(s(-1))\\
      &=V(1)
    \end{aligned}
  \end{equation*}
        Hence $w=0$ is not a minimiser of $V(w)$. 
        If $\sigma > 1$, from conditions~\ref{con2} and \ref{con44}, we get
        \begin{equation*}
    \begin{aligned}
        \frac{dV(w)}{dw}|_{w=0}&=p(x)g'(s(0))s'(0)-(1-p(x))g'(s(0))s'(0)\\
        &=(2p(x)-1)g'(s(0))s'(0)\\
        &\neq 0.
    \end{aligned}
        \end{equation*}
        Hence, $w=0$ is not a minimiser of $V(w)$. 
        Therefore, we obtain $w^\ast > 0$ if $p(x) > 0.5$ and $w^\ast < 0$ otherwise.
        In conclusion, $\text{sign}(w^\ast)=\text{sign}(p-\frac{1}{2})$.  
\end{enumerate}
\qed

\noindent{\textbf{Proof of Theorem~\ref{thm:conv1}}}
\begin{enumerate}[(i)]
	\item
Denote $h(z)=-g(z), \varphi(v)$ the conjugate function of $h(z)$ defined by $\varphi(v)=\sup_z(vz-h(z))$. Suppose that $vz-h(z)$ attains its maximum at $z^\ast$ for fixed $v$, then $p(z^\ast)=-vz^\ast+h(z^\ast)$ attains its minimum. We have $0 \in \partial p(z^\ast)=-v + \partial h(z^\ast)$ or $v \in \partial h(z^\ast)$, and
\begin{equation}\label{eqn:pro3}
    \varphi(v)=vz^\ast-h(z^\ast).
\end{equation}
In convex analysis, the converse holds.  
 Denote $\varphi^\ast(z)$ the conjugate of $\varphi(v)$. Namely,
 \begin{equation}\label{eqn:pro33}
 \varphi^\ast(z)=\sup_v(vz-\varphi(v)).
 \end{equation}
 Suppose that $vz-\varphi(v)$ attains its maximum at $v^\ast$ for fixed $z$, then $q(v^\ast)=-v^\ast z+\varphi(v^\ast)$ attains its minimum. Hence, we obtain $z \in \partial\varphi(v^\ast)$ and
\begin{equation}\label{eqn:pro4}
    \varphi^\ast(z)=v^\ast z-\varphi(v^\ast).
\end{equation}
Again, the converse holds since $\varphi(v)$ is convex.
With $h(z)$ closed, the conjugate of conjugate function recovers \citep[Proposition 3.4.2]{lange2016mm}, i.e., 
\begin{equation}\label{eqn:pro6}
    \varphi^\ast(z)=h(z).
\end{equation}
Together with (\ref{eqn:pro3}) and (\ref{eqn:pro4}), $v \in \partial h(z^\ast)$ is equivalent to $z\in \varphi(v^\ast)$. Furthermore, from (\ref{eqn:pro33})-(\ref{eqn:pro6}) we have
\begin{equation*}
    h(z) \geq vz-\varphi(v);\ h(z)=v^\ast z - \varphi(v^\ast),
\end{equation*}
which is the same as
\begin{equation*}
    g(z) \leq -vz+\varphi(v); \ g(z)=-v^\ast z + \varphi(v^\ast).
\end{equation*}
Thus $-vz+\varphi(v)$ majorises $g(z)$ at $v^\ast$.
In Algorithm~\ref{alg:pcoco},  given $z_i$, if $v_i \in \partial (-g(z_i))$ or $z_i \in \partial \varphi(v_i)$, then $-v z_i+\varphi(v)$ is minimised with respect to $v$. 
With Step 3-5 in Algorithm~\ref{alg:pcoco}, $z_i=s(u_i(\bm\beta^{(k)}))$, we get 
\begin{equation}\label{eqn:mm6}
	F(\bm\beta^{(k+1)}) \leq Q(\bm\beta^{(k+1)}|\bm\beta^{(k)}) \leq Q(\bm\beta^{(k)}|\bm\beta^{(k)}) 
	=F(\bm\beta^{(k)}), 
\end{equation}
where the surrogate loss is given by
\begin{equation*}
    \begin{aligned}
    Q(\bm\beta|\bm\beta^{(k)})&=\sum_{i=1}^ns(u_i(\bm\beta))\left(-v_i^{(k+1)}(\bm\beta^{(k)})\right)+\varphi\left(v_i^{(k+1)}(\bm\beta^{(k)})\right)+\Lambda(\bm\beta)\\ 
    &=\ell(\bm\beta|\bm\beta^{(k)})+\Lambda(\bm\beta).
    \end{aligned}
\end{equation*}
To minimise $Q(\bm\beta|\bm\beta^{(k)})$ in Step 5, the objective function is simplified since $v_i^{(k+1)(\bm\beta^{(k)})}$ is a constant in the current iteration step:
\begin{equation*}\label{eqn:pro9}
	\begin{aligned}
        \argmin_{\bm\beta} Q(\bm\beta|\bm\beta^{(k)})&=\argmin_{\bm\beta} \sum_{i=1}^ns(u_i(\bm\beta))\left(-v_i^{(k+1)}(\bm\beta^{(k)})\right)+\varphi\left(v_i^{(k+1)}(\bm\beta^{(k)})\right)+\Lambda(\bm\beta)\\ 
                                                     &=\argmin_{\bm\beta} \sum_{i=1}^n s(u_i(\bm\beta))\left(-v_i^{(k+1)}(\bm\beta^{(k)})\right)+\Lambda(\bm\beta). 
	\end{aligned}
\end{equation*}
Furthermore, by assumption $g(z)$ is bounded below, hence for every $z, g(z) \geq c$ for some constant $c$. From (\ref{eqn:mloss2}), (\ref{eqn:plik}) and $\Lambda(\bm\beta) \geq 0$, we get $F(\bm\beta^{(k)})\geq c$. In summary, the sequence $F(\bm\beta^{(k)})$ is nonincreasing and bounded below. Hence the sequence $F(\bm\beta^{(k)})$ of Algorithm~\ref{alg:pcoco}
converges.
	\item From (\ref{eqn:mm6}), $Q(\bm\beta|\bm\beta^{(k)})$ majorises $F(\bm\beta)$ at $\bm\beta^{(k)}$. Since $g$ and $s$ are differentiable, $L(\bm\beta)$ and $\ell(\bm\beta|\bm\beta^{(k)})$ are differentiable with respect to $\bm\beta$.
		Furthermore, since $s(u)(-v)+\varphi(v)$ is jointly continuous in $(u, v)$, 
		$\ell(\bm\beta|\bm\beta^{(k)})$ is jointly continuous in $(\bm\beta, \bm\beta^{(k)})$. Applying Theorem 7 in \citet{wang2022mm}, we obtain the desired results provided that the penalty function 
		$p_\lambda(|\beta_j|)$ 
satisfies the following assumptions:
\begin{assumption}\label{ass:pen}
  $p_\lambda(\theta)$ is continuously differentiable, nondecreasing and concave on $(0, \infty)$ with $p_\lambda(0)=0$ and $0 < p'_\lambda(0+) < \infty$.
\end{assumption}
		\end{enumerate}
		\qed
\putbib[wangres]
\end{bibunit}
\end{appendices}
\end{document}